\documentclass[12pt]{article}

\onecolumn 

\usepackage{enumitem}
\usepackage{graphicx}
\usepackage{amsmath,amsthm,amssymb, xcolor}
\usepackage[round]{natbib}
\usepackage[a4paper, total={7.5in, 10.5in}]{geometry}
\usepackage{longtable}
\usepackage{array}

\newtheorem{remark}{Remark}
\newtheorem{theorem}{Theorem}
\newtheorem{assumption}{Assumption}
\newtheorem{example}{Example}

\newcommand{\bs}{\boldsymbol}

\begin{document}

\begin{center}

\Large
Estimating conditional Mann-Whitney effects \\ using pseudo-observation-based regression
\\[0.5cm]
\large
\textbf{Dennis Dobler}\footnote{ORCID: 0000-0002-9040-0854}
\normalsize
\\[0.2cm]
Institute of Statistics and Mathematical Economics, \\
RWTH Aachen University,
\\
Kreuzherrenstr.\ 2,
\\
D-52062 Aachen, Germany
\\
Corresponding author. {email: dennis.dobler@rwth-aachen.de}
\\[0.4cm]
\large
\textbf{Alina Schenk}\footnote{ORCID: 0000-0003-1998-2625}
\normalsize
\\[0.2cm]
Institute for Medical Biometry, Informatics and Epidemiology, \\
University of Bonn, \\
University Hospital Bonn, \\
Venusberg-Campus 1,
\\
D-53127 Bonn, Germany
\\[0.4cm]
\large
\textbf{Matthias Schmid}\footnote{ORCID: 0000-0002-0788-0317}
\normalsize
\\[0.2cm]
Institute for Medical Biometry, Informatics and Epidemiology, \\
University of Bonn, \\
University Hospital Bonn, \\
Venusberg-Campus 1,
\\
D-53127 Bonn, Germany
\\[0.3cm]
April 1, 2026

\end{center}



\abstract{\normalfont \normalsize
\noindent The Mann-Whitney effect is an effect measure for the order of two sample-specific outcome variables.
    It has the interpretation of a probability and also a connection to the area under the ROC curve.
    In the literature it has been considered for both ordinal and right-censored time-to-event outcomes.
    For both cases, the present paper introduces a distribution-free regression model that relates the Mann-Whitney effect to a linear combination of covariates. To fit the model, we develop a pseudo-observation-based procedure yielding consistent and asymptotically normal coefficient estimates. In addition, we propose bootstrap-based hypothesis tests to infer the effects of the covariates on the Mann-Whitney effect. A simulation study on the small-sample behavior of the proposed method demonstrates that the novel hypothesis tests keep up with the z-test of a Cox regression model. The new methods are used to analyze progression-free survival in breast cancer patients enrolled for the randomized phase~III SUCCESS-A trial.} 

    \vspace{0.2cm}








\section{Introduction}
\label{sec:intro}

The two-sample problem has a long-standing tradition in biostatistical research.
There exist numerous methods for testing the difference between two distributions or aspects thereof, including the two-sample $t$-test, Wilcoxon signed-rank test, Mann-Whitney $U$ test, and logistic regression.
In the context of time-to-event data, a large number of  approaches for comparing two survival functions is available as well. For example, as an extension of the classical log-rank test, weighted log-rank tests \citep{difri20, royston2020}, accelerated failure time models, or the Cox proportional hazards model \citep{cox72} could be used. If one treats the treatment group indicator as a covariate, it is also possible to employ flexible approaches involving inverse-probability-of-censoring weighting \citep{blahoschei23, ov25} or the jackknife-based pseudo-observation approach for regression \citep{anklero03}; here the idea is to solve a generalized estimating equation based on re-weighted individual observations and pseudo-observations, respectively, to find an estimator of the parameter quantifying the group effect.

In this paper, we take a somewhat different approach: We focus on a statistical estimand with an appealing interpretation that intrinsically incorporates the two-sample problem. More specifically, we analyze the {\em Mann-Whitney effect}, sometimes called {\em relative (treatment) effect}, $\theta = P(T_1 > T_2)$, where $T_1, T_2$ describe the event times of two random and independent individuals from sample groups $j=1,2$.
Often the term $0.5 \cdot P(T_1 = T_2)$ is added to $\theta$ as a continuity correction to account for ties.
Thus, under the null hypothesis of no group differences, it holds that $\theta = 0.5$ if the distribution functions are continuous. As shown, e.g., by \cite{dopa18}, approaches based on~$\theta$ offer easy-to-interpret alternatives to log-rank tests in two-sample comparisons. Building on this methodology, we propose a jackknife-based pseudo-observation technique that enables adjusting $\theta$ for covariates, thereby allowing for the analysis of both randomized and observational data within an integrated framework.\\

\noindent {\em Motivation of the proposed method.} The development of the proposed method has been motivated by our re-analysis of the multicenter randomized phase III SUCCESS-A study (\citealt{GregorioHaeberleFaschingetal.2020, schenk2025}). Between 2005 and 2007, SUCCESS-A enrolled a total of 3,754 female patients with primary invasive breast cancer, assigning study participants randomly to one of two treatment arms (control: standard adjuvant chemotherapy, intervention: standard adjuvant chemotherapy with the addition of gemcitabine). The primary outcome of the study was disease-free survival (DFS), defined as the period from the date of randomization to the earliest date of disease progression (distant metastases, local
and contra local recurrence, and secondary primary tumors) or death from any cause. For the analysis of the primary outcome, \cite{GregorioHaeberleFaschingetal.2020} used a univariable (``marginal'') Cox regression model with treatment as the only covariate. This model did not show any evidence for improved DFS when adding gemcitabine to standard chemotherapy (DFS hazard ratio = 0.93 [intervention vs.\@ control]; 95\% confidence interval [0.78;1.12]). For secondary analysis, the authors used a multivariable Cox regression including main and interaction effects between treatment and covariates measured at baseline, investigating whether variables like age, body mass index and tumor stage are predictive for the effect of treatment effect DFS. Although not being part of the confirmatory analysis of the SUCCESS-A study, this model revealed several interesting patterns in subgroups of the study population (e.g., variations of treatment effects by tumor type and tumor stage).

While the Cox-based strategy by \cite{GregorioHaeberleFaschingetal.2020} constitutes an established analysis approach in clinical trial research, it naturally raises the question whether subgroup two-sample comparisons could alternatively be performed using a {\em distribution-free} statistical model. In particular, such a model could help assessing the probability of whether a specific patient will benefit from the intervention without making potentially restrictive assumptions like the proportional hazards assumption. As we will show below, our proposed model is able to address this question within a unified theoretical framework, specifying $T_1$ and $T_2$ as the survival times in the intervention and control groups, respectively, and adjusting the Mann-Whitney effect~$\theta$ by the values of patient-specific covariates ${\bs Z_1}, {\bs Z_2}$.\\

\noindent {\it Related work.} Variants of the Mann-Whitney effect $\theta$ enjoy popularity in many statistical applications and are subject to extensive methodological research. In the uncensored case, \cite{brumu00} developed methodology for a tie-adjusted variant of $\theta$; optimal sample sizes were derived by \cite{hababru19}. Recently, \cite{bruko25} developed an unbiased variance estimator with the help of rank-based methods. \cite{schukobru25} conducted a comparison of multiple tests based on Mann-Whitney effects, including a novel test with improved variance estimation.
Recently, \cite{thisabazi25} analyzed a nonparametric covariate adjustment to the Mann-Whitney effect which they termed ``NANCOVA''.
A comprehensive nonparametric treatment of Mann-Whitney effects, including factorial designs, is given in the book by \cite{brubako18}.

In the context of randomly right-censored data, \cite{efron67} was the first to propose an estimator for $\theta$.
\cite{emdidomu24} extended the nonparametric multiple-sample comparison approach within general factorial designs \citep{dopa20} to dependent censoring models using the copula-graphic estimator.

In addition to these works, there have been several approaches for establishing semi-parametric extensions of Mann-Whitney effects under fully observed data. For example, \cite{brupeal06}, \cite{zhazhatu11}, and \cite{razh12} proposed generalized linear regression models for the area under the ROC curve (AUC) which is related to the Mann-Whitney effect. In bioinformatics, \cite{mahuang2005} and \cite{mayrschmid2014} proposed regularized AUC regression methods for binary and time-to-event outcomes, respectively. Machine learning approaches to AUC optimization were discussed by \cite{yayi22}. A related methodological approach is the probabilistic index model \citep{thanecle12, netha15}. In contrast to the method proposed here, probabilistic index models  incorporate the effect of covariate {\em differences} ${\bs Z_1}-{\bs Z_2}$ but not of the actual covariate values ${\bs Z_1}, {\bs Z_2}$ in groups $j=1,2$. \\

\noindent {\it Objective and structure of this paper.}
The present paper is the first to propose a regression approach for the Mann-Whitney effect allowing for randomly right-censored and possibly tied data.  
It is based on a generalized linear regression model for~$\theta$ where the effects of the involved covariates ${\bs Z_1},{\bs Z_2}$ have a natural interpretation as group-covariate interactions. Using a two-sample jackknife pseudo-observation approach for model fitting, we prove a central limit theorem for the parameter estimators in both the uncensored and censored cases.

The rest of this paper is organized as follows. Section~\ref{ssec:uncensored} offers a gentle description of the novel approach when all data are fully observed. Section~\ref{sec:cens} extends the methodology to the censored case. Section~\ref{sec:simus} presents the results of a comprehensive simulation study that compares the proposed method to a Cox-based approach with group-covariate interactions.
Furthermore, we propose and apply bootstrap-based
hypothesis tests to infer the effects of the covariates on the Mann-Whitney effect. In Section~\ref{sec:data} we  analyze DFS in the SUCCESS-A study data, demonstrating how the proposed method can be used to identify patients with a high probability of prolonged survival in the intervention group. We conclude with a discussion and aspects for future research in Section~\ref{sec:disc}.
The supplementary material contains all proofs, additional mathematical considerations, additional simulation results, and supplementary results from our data analyses.

\section{Pseudo-observation-based regression for two-sample problems}
\label{sec:pseudo}

\vspace*{.1cm}
\subsection{Fully observable case}
\label{ssec:uncensored}

For each sample group $j=1,2$, let $(T_{ji_j}, \bs Z_{ji_j}^\top)^\top, i_j=1, \dots, n_j $, be independent and identically distributed (i.i.d.) random vectors on a probability space $(\Omega, \mathcal A, P)$ with  $T_{11}, \dots, T_{1n_1} \sim S_1$ and $T_{21}, \dots, T_{2n_2} \sim S_2$. In view of the extension to censored observations in Section~\ref{sec:cens}, we describe the model and method in terms of the survival functions $S_1(t) = P(T_{1i_1} > t)$ and $S_2(t) = P(T_{2i_2} > t), t \in \mathbb R$; alternative formulations in terms of cumulative distribution functions are straightforward. For now, we consider the \textit{uncensored}, i.e., fully observable case without censoring.
In fact, the event times $T_{ji_j}$ only need to be ordinally scaled and not necessarily continuous random variables. The $T_{ji_j}$'s are real-valued, albeit not necessarily strictly positive in this section yet; the covariates $\bs Z_{ji_j}$'s are $\mathbb R^{p_j}$-valued random variables, $p_j \in \mathbb N$, possibly including sets of dummy variables representing factor covariates. Note that the first entries of the $\bs Z_{ji_j}$'s are not necessarily all equal to 1 because the intercept term will be modeled separately.

We propose and fit a model for the conditional probability
\begin{align}
\label{eq:model1}
 \theta(\bs z_1, \bs z_2) = P(T_{1i_1} > T_{2i_2} \, | \, \bs Z_{1i_1}=\bs z_1, \bs Z_{2i_2} = \bs z_2).
\end{align}
In principle, both covariate vectors could address completely different characteristics (or features) of an individual.
In most applications, however, they will refer to the same set of characteristics.
One interesting special case is $\bs z_1 = \bs z_2 =\bs  z$.
For example, if both sample groups are defined by different treatment regimes, then $\theta(\bs z) = \theta(\bs z,\bs z)$ quantifies the probability that some individual $i_1$ with covariate values $\bs z$ under Treatment~1 would live longer than some individual $i_2$ with the same covariate values $\bs z$ under Treatment~2. Considering this can allow for drawing causal inferences under correct model specification in the absence of other confounding (see Section~\ref{sec:disc}).

In general, we allow both dimensions, $p_1$ and $p_2$, to differ. For example, this could be useful if the first sample refers to data from a previous study with few covariates and the second sample was more recently collected, with additional covariates.
Then $\theta(\bs z^{(1)},(\bs z^{(1)\top},\bs z^{(2)\top})^\top )$ quantifies the Mann-Whitney effect of individuals for which only the covariates $\bs z^{(1)}$ are taken into account for the first group's event time but in addition also $\bs z^{(2)}$ for the second group's event time. Alternatively, such a scenario could be interpreted as a comparison with a subpopulation from group~2.

We assume that the estimand $\theta$ follows a model that can be described with the help of a strictly monotone inverse link function $\mu: \mathbb R \to \mathbb R$ and a parameter vector $\bs \beta \in \mathbb R^{1+p_1+p_2}$, where $\bs \beta^\top = (\beta_0, \bs \beta_1^\top, \bs \beta_2^\top)^\top$. Then the model is defined as
\begin{align}
\label{eq:model2}
 \theta(\bs z_1,\bs  z_2) = \mu(\beta_0 + \bs \beta_1^\top \bs z_1 + \bs \beta_2^\top\bs  z_2).
\end{align}
If needed, the model could of course be reduced: {\color{black}For instance, if the value of the intercept term is known, $\beta_0 = \check \beta_0$, the model could be reduced to $\theta(\bs z_1,\bs  z_2) = \check \mu(\bs \beta_1^\top \bs z_1 + \bs \beta_2^\top\bs  z_2)$ where $\check \mu( \ \cdot \ ) = \mu(\check \beta_0 + \ \cdot \ ) $.}
{\color{black}We assume in the following that  model~\eqref{eq:model2}  describes the reality for a problem at hand, i.e.\@ for a ``true'' parameter vector $\bs \beta_0 \in \mathbb R^{1+p_1+p_2}$ printed in bold-type, not to be confused with the intercept term $\beta_0 \in \mathbb{R}$.}

A nonparametric estimator for $\theta$ is given by
\begin{align}
\label{eq:nonparametric_est}
 \hat \theta = \hat P(T_{1i_1} > T_{2i_2}) = \frac1{n_1}\frac1{n_2} \sum_{i_1=1}^{n_1} \sum_{i_2=1}^{n_2} 1\{T_{1i_1} > T_{2i_2}\} = \frac1{n_2} \sum_{i_2=1}^{n_2}\hat S_1(T_{2i_2}) = 1- \frac1{n_1}\sum_{i_1=1}^{n_1} \hat S_2(T_{1i_1}-),
 \end{align}
 where $1\{ \cdot \}$ denotes the indicator function of an event, $\hat S_j(t) = \frac1{n_j} \sum_{i_j=1}^{n_j} 1\{T_{j i_j} > t\}$, $j=1,2$, denotes the empirical survival function, and $\hat S_2(t-)$ denotes the left-hand limit of $\hat S_2$ at $t\in \mathbb R$.
 Later $\hat S_j$ will be replaced by the Kaplan-Meier estimator based on randomly right-censored observations.

Motivated by Eq.\@ \eqref{eq:nonparametric_est}, we define \textit{two-sample pseudo-observations} by
\begin{align}
\label{eq:pv}
\begin{split}
 \tilde \theta_{i_1 i_2} & := - \int_0^\infty \tilde S_{1i_1}(u) \, d \tilde S_{2i_2}(u) \\
 & := - \int_0^\infty (n_1 \hat S_1 - (n_1-1) \hat S_1^{(i_1)})(u) \, d (n_2 \hat S_2 - (n_2-1) \hat S_2^{(i_2)})(u) \\
 & = - n_1 n_2\int_0^\infty  \hat S_1(u) \, d \hat S_2(u)  
    + (n_1-1)n_2 \int_0^\infty \hat S_1^{(i_1)}(u) \, d\hat S_2(u)
    + n_1(n_2-1) \int_0^\infty \hat S_1(u) \, d \hat S_2^{(i_2)}(u) \\
    & \qquad 
    - (n_1-1)(n_2-1) \int_0^\infty \hat S_1^{(i_1)}(u) \, d \hat S_2^{(i_2)}(u)\\
 & =: n_1 n_2 \,\hat \theta - (n_1-1)n_2\, \hat \theta_1^{(i_1)} - n_1 (n_2-1) \,\hat \theta_2^{(i_2)} + (n_1-1)(n_2-1)\, \hat \theta_{12}^{(i_1 i_2)} .
 \end{split}
\end{align}
The integrals are to be understood in the Riemann-Stieltjes manner.
Here, $\hat \theta$ is the estimator defined in \eqref{eq:nonparametric_est}, $\hat \theta_1^{(i_1)}$ is the same estimator (with factor $((n_1-1)n_2)^{-1}$ but based on all random variables except the $i_1$-th from sample group $1$, and $\hat \theta_2^{(i_2)}$ is defined analogously. The estimator $\hat \theta_{12}^{(i_1 i_2)} $ (with factor $(n_1-1)^{-1}(n_{2}-1)^{-1}$) omits the $i_1$-th random variable from group~1 and the $i_2$-th random variable from group~2.
A similar notation has been used for the reduced estimators $\hat S_j^{(i_j)}$, $j\in \{1,2\}$.
In the fully observable case, it is easy to see that $\tilde \theta_{i_1 i_2}$ reduces to $1\{T_{1i_1} > T_{2i_2}\}$, which has expectation $\text{E}(\tilde \theta_{i_1 i_2}) = P(T_{1i_1} > T_{2i_2}) =: \theta.$
Definition~\eqref{eq:pv} is also feasible in the presence of right-censoring; see Section~\ref{sec:cens} below.

\begin{remark}
The formulas in~\eqref{eq:nonparametric_est} and~\eqref{eq:pv} naturally allow for a continuity correction in the case of ties: If $\hat \theta$ is defined by $\hat \theta = \sum_{i_1=1}^{n_1} \sum_{i_2=1}^{n_2} ( 1\{T_{1i_1} > T_{2i_2}\} + 0.5 \cdot 1\{T_{1i_1} = T_{2i_2}\} )/ (n_1 n_2)$, it is easy to see that $\tilde \theta_{i_1 i_2} = 1\{T_{1i_1} > T_{2i_2}\} + 0.5 \cdot 1\{T_{1i_1} = T_{2i_2}\} $.
\end{remark}

\begin{remark}
The definition in~\eqref{eq:pv} seems to be in contrast to \citet[p.~15]{efron79} who stated that not pairs of observations should be left out but rather all observations of both samples consecutively, without such pairings.
Efron's arguments are based on two related papers \citep{miller74, hinkley77} which both considered linear regression problems.
Some authors indeed apply the one-sample jackknife to both samples consecutively, also in censored data problems; see e.g.\ \cite{chang12}.
On the other hand, \cite{schewa04} proposed the two-sample jackknife as applied in~\eqref{eq:pv}. \cite{chuang12} applied this type of two-sample jackknife method to the Jaccard index.
\end{remark}

In order to relate $\theta$ to the covariate values $\bs  Z_{1i_1}, \bs Z_{2i_2}$ in accordance with model \eqref{eq:model2}, we fit a regression model by solving the following generalized estimating equation (GEE), {\color{black}similar to the approach by \cite{razh12} for completely observable outcomes}:
\begin{align}
 \label{eq:gee}
 \bs U(\bs \beta) = \sum_{i_1=1}^{n_1} \sum_{i_2=1}^{n_2} \bs A_{i_1 i_2}(\tilde \theta_{i_1 i_2} - \mu_{i_1 i_2} ) = \sum_{i_1,i_2} \bs U_{i_1 i_2}(\bs \beta) = \bs 0,
\end{align}
where
$$ \bs A_{i_1 i_2} = A(\bs \beta, \bs Z_{1i_1}, \bs Z_{2i_2}) = (\nabla_{\bs \beta} \mu_{i_1 i_2} )^\top V_{i_1 i_2}^{-1} = (1, \bs Z_{1i_1}^\top, \bs Z_{2i_2}^\top)^\top \, \mu'((1, \bs Z_{1i_1}^\top, \bs Z_{2i_2}^\top) \bs \beta) \, V_{i_1 i_2}^{-1}, $$
$V_{i_1 i_2}$ is a working (co)variance (matrix) for $\tilde \theta_{i_1 i_2}$, and
$ \mu_{i_1 i_2} = \mu (\beta_0 + \bs \beta_1^\top \bs Z_{1i_1} + \bs \beta_2^\top \bs Z_{2i_2} ). $
This approach is motivated by the one-sample-based equation in \cite{anklero03}; see their Equation~(3$\cdot$1).
In the present case, the estimand $\theta$ is univariate, so $V_{i_1i_2}$ can be set equal to 1.

Note that Condition~2 in \cite{anklero03} postulates the independence of the terms $\bs U_i(\bs \beta)$ contributing to their GEE.
They also argue that the pseudo-observations (in the one-sample case) are asymptotically independent.
While this is certainly so, the correlation structure between the pseudo-observations still plays a significant role in the asymptotic behavior of the solution $\hat{\bs \beta}$ of \eqref{eq:gee}; we  refer to \cite{ovpape17} who showed that the aforementioned correlation structure affects the asymptotic covariance matrix of the normalized parameter estimator in the one-sample and randomly right-censored case. In our two-sample case, 
$\tilde \theta_{i_1 i_2}$ and $\tilde \theta_{\tilde i_1 \tilde i_2}$ are generally dependent if either $i_1 = \tilde i_1$ and/or $i_2 = \tilde i_2$.


\begin{example}[Identity link function]
\label{ex:identity}

In the case of the univariate identity link function, i.e.,\@ $\mu(x) = x$,
we have
$$ A(\bs \beta,\bs z_1,\bs z_2) = (1, \bs z_1^\top, \bs z_2^\top )^\top.$$
Hence the GEE  \eqref{eq:gee} takes the form
$$ \bs U(\bs \beta) = \sum_{i_1=1}^{n_1} \sum_{i_2=1}^{n_2}
    \begin{pmatrix} 1 \\\bs  Z_{1i_1} \\ \bs Z_{2i_2} \end{pmatrix}
    \Big(\tilde \theta_{i_1i_2} - \beta_0 - \bs \beta_1^\top \bs Z_{1i_1} - \bs \beta_2^\top \bs Z_{2i_2} \Big) = \bs 0, $$
which is solved by
$$  \hat{\bs \beta} := \hat{\bs \Sigma}^{-1} \hat {\bs \Psi} := \Big( \frac{1}{n_1 n_2} \sum_{i_1=1}^{n_1} \sum_{i_2=1}^{n_2} \Big(\begin{smallmatrix} 1 \\ \bs Z_{1i_1} \\\bs  Z_{2i_2} \end{smallmatrix}\Big)\Big(\begin{smallmatrix} 1 \\ \bs Z_{1i_1} \\\bs  Z_{2i_2} \end{smallmatrix} \Big)^\top \Big)^{-1} \frac{1}{n_1 n_2}  \sum_{i_1=1}^{n_1} \sum_{i_2=1}^{n_2}  \Big(\begin{smallmatrix} 1 \\ \bs Z_{1i_1} \\\bs  Z_{2i_2} \end{smallmatrix} \Big) \tilde \theta_{i_1 i_2}.$$
Here, the inverse $\hat{\bs \Sigma}^{-1}$ is to be understood as the generalized Moore-Penrose pseudo-inverse of $\hat{\bs \Sigma}$.
If the second moments of all covariates exist,  $\hat{\bs \Sigma}$ converges almost surely to
$$ \bs \Sigma :=  \text{E} \Bigg(\begin{pmatrix} 1 \\ \bs Z_{1i_1} \\ \bs Z_{2i_2} \end{pmatrix}\begin{pmatrix} 1 \\ \bs Z_{1i_1} \\ \bs Z_{2i_2} \end{pmatrix}^\top \Bigg) =  \begin{pmatrix}
    1 & \bs E_{1}^\top & \bs E_2^\top \\
    \bs E_1 & \bs \Sigma_1 + \bs E_1 \bs E_1^\top & \bs E_1 \bs E_2^\top \\
    \bs E_2 & \bs E_2 \bs E_1^\top & \bs \Sigma_2 + \bs E_2 \bs E_2^\top
\end{pmatrix}$$
with expectation vectors $\bs E_j = \text{E}(\bs Z_{j i_j}) \in \mathbb R^{p_j}$ and variance-covariance matrices $\bs \Sigma_j = \text{var}(\bs Z_{j i_j}) \in \mathbb R^{p_j \times p_j}$, $j=1,2$.
It is easy to see that $\bs \Sigma$ has full rank $p = 1 + p_1 + p_2$ whenever $\bs \Sigma_1$ and $\bs \Sigma_2$ have full ranks $p_1$ and $p_2$, respectively:
First, being the limit of a sum of positive semi-definite matrices, $\bs \Sigma$ is positive semi-definite as well; second, for all $\bs x = (x_0, \bs x_1^\top , \bs x_2^\top)^\top \in \mathbb R^{1+p_1+p_2} \setminus \{\bs 0\}$, and using the stochastic independence of $\bs Z_{1i_1} $ and $\bs Z_{2i_2}$, it holds that
\begin{align*}
   \bs x^\top \bs \Sigma \bs x
    & = \bs x^\top \text{E}\Big(\Big(\begin{smallmatrix}
    1 \\ \bs Z_{1i_1} \\ \bs Z_{2i_2} 
\end{smallmatrix} \Big)\Big(\begin{smallmatrix}
    1 \\ \bs Z_{1i_1} \\ \bs Z_{2i_2} 
\end{smallmatrix} \Big)^\top \Big) \bs x 
= \bs x^\top \Big[ \text{var}\Big(\begin{smallmatrix}
    1 \\ \bs Z_{1i_1} \\ \bs Z_{2i_2} 
\end{smallmatrix} \Big)
 +  \Big( \begin{smallmatrix}
     1 \\ \bs E_1 \\ \bs E_2
 \end{smallmatrix} \Big) \Big( \begin{smallmatrix}
     1 \\ \bs E_1 \\ \bs E_2
 \end{smallmatrix} \Big)^\top \Big] \bs x 
    \\
    & = \text{var} \Big(\bs x^\top \Big(\begin{smallmatrix}
    1 \\ \bs Z_{1i_1} \\ \bs Z_{2i_2} 
\end{smallmatrix} \Big) \Big) + \big((1, \bs E_1^\top, \bs E_2^\top)\bs x \big)^2
 = (x_0 + \bs x_1^\top \bs E_1 + \bs x_2^\top \bs E_2)^2 + \text{var}(\bs x_1^\top \bs Z_{1i_1}) + \text{var}(\bs x_2^\top \bs Z_{2i_2})\\
 & = (x_0 + \bs x_1^\top \bs E_1 + \bs x_2^\top \bs E_2)^2 + \bs x_1^\top \bs \Sigma_1 \bs x_1 + \bs x_2^\top \bs \Sigma_2 \bs x_2 > 0
\end{align*}
if $\bs \Sigma_1$ and $\bs \Sigma_2$ have full ranks.
Hence, $\bs x^\top \bs \Sigma \bs x$ equals zero only if $\bs x_j^\top \bs \Sigma_j \bs x_j = 0$, $j=1,2$. This is equivalent to $\bs x_1 = \bs x_2 = \bs 0$, implying $x_0 = 0$.
Consequently, $\bs \Sigma$ is positive definite and the probability that the  inverse of $\hat{\bs \Sigma}$ exists tends to one with increasing sample sizes.

At present, i.e., in the fully observable case, the parameter estimator reduces to
\begin{align}
    \label{eq:beta_hat_reduced}
 \hat{\bs \beta} = \hat {\bs \Sigma}^{-1} 
\Big(\hat \theta \ , \ \frac1{n_1} \sum_{i_1=1}^{n_1}\bs  Z_{1i_1}^\top \tilde \theta_{i_1 \bullet} \ , \  \frac1{n_2} \sum_{i_2=1}^{n_2} \bs Z_{2i_2}^\top \tilde \theta_{\bullet i_2}\Big)^\top
\end{align}
since $\hat \theta =  \sum_{i_1=1}^{n_1} \sum_{i_2=1}^{n_2} \tilde \theta_{i_1 i_2} / (n_1 n_2)$.
Here we used the definitions
\begin{align*}
\tilde \theta_{i_1 \bullet} & := \frac1{n_2} \sum_{i_2=1}^{n_2} \tilde \theta_{i_1 i_2} = \frac1{n_2} \sum_{i_2=1}^{n_2} 1\{T_{1i_1}> T_{2i_2}\} = 1 -\hat S_2(T_{1 i_1}-) \\\text{and} \quad \tilde \theta_{\bullet i_2} & :=  \frac{1}{n_1} \sum_{i_1=1}^{n_1} \tilde \theta_{i_1 i_2} = \frac1{n_1} \sum_{i_1=1}^{n_1} 1\{T_{1i_1}> T_{2i_2}\} = \hat S_1(T_{2 i_2}) .
\end{align*}

To analyze the asymptotic properties of $\hat{\bs \beta}$, we define 
\begin{align*}
    \bs \Omega^{(1)} \ := \   & \text{cov}\Big(\Big(\begin{smallmatrix} 1 \\ \bs Z_{1i_1} \\ \bs Z_{2i_2} \end{smallmatrix}\Big) 1\{T_{1i_1} > T_{2i_2} \}, \Big(\begin{smallmatrix} 1 \\ \bs Z_{1i_1} \\ \bs Z_{2\tilde{i}_2} \end{smallmatrix}\Big) 1\{T_{1i_1} > T_{2\tilde{i}_2} \} \Big),  \quad i_2 \neq \tilde i_2, \\
    \text{and} \quad 
\bs \Omega^{(2)} \ := \   &\text{cov}\Big(\Big(\begin{smallmatrix} 1 \\ \bs Z_{1i_1} \\ \bs Z_{2i_2} \end{smallmatrix}\Big) 1\{T_{1i_1} > T_{2i_2} \}, \Big(\begin{smallmatrix} 1 \\ \bs Z_{1\tilde{i}_1} \\ \bs Z_{2i_2} \end{smallmatrix}\Big) 1\{T_{1\tilde{i}_1} > T_{2i_2} \} \Big), \quad i_1 \neq \tilde i_1.
\end{align*}
Assuming that $n_1/(n_1+n_2) \to \lambda \in (0,1)$ as $\min(n_1,n_2) \to \infty$, that $\text{E}(\bs Z_{j i_j}^\top \bs Z_{j i_j}) < \infty$, $j=1,2$,  and that both $\bs \Omega := (1-\lambda) \bs \Omega^{(1)} + \lambda \bs \Omega^{(2)}$ and $\bs \Sigma$ have full rank $p$,
then, as $\min(n_1,n_2) \to \infty$,
$$ \sqrt{\frac{n_1 n_2}{n_1 + n_2}} \, (\hat {\bs \beta} - \bs \beta_0)  \, \stackrel d \to \, \mathcal N_p(\bs 0, \bs \Sigma^{-1}\bs  \Omega\bs  (\bs \Sigma^{-1})^\top). $$
The proof of this central limit theorem is given in Appendix \ref{app:proofs}.

In practice, this result could be used to obtain confidence regions for $\bs \beta_0$ after employing an estimator of the asymptotic variance-covariance matrix.
A consistent estimator of $\bs \Sigma$ was suggested above; a consistent estimator of $\bs \Omega$ is proposed in Appendix \ref{app:proofs}.
\end{example}

\subsection{The censored case}
\label{sec:cens}

In this section, we assume that the random variables of interest, $T_{ji_j}$, are {\color{black}randomly right-censored.}
That is, the data are presented by sample-wise i.i.d.\ random variables
$$  (X_{11}, \delta_{11}, \bs Z_{11}^\top)^\top, \dots, (X_{1n_1}, \delta_{1n_1}, \bs Z_{1n_1}^\top)^\top 
\quad \text{and} \quad (X_{21}, \delta_{21}, \bs Z_{21}^\top)^\top, \dots, (X_{2n_2}, \delta_{2n_2}, \bs Z_{2n_2}^\top)^\top , $$
where $X_{j i_j} = \min(T_{j i_j}, C_{j i_j})$ and $\delta_{j i_j} = 1\{ T_{j i_j} \leq C_{j i_j}\}$ denote the observed (possibly right-censored) event times and the status indicators, respectively, for individual $i_j \in \{1,\dots, n_j \}$ in group $j =1,2$. The censoring times $C_{j i_j}$ are assumed to be stochastically independent of $(T_{j i_j}, \bs Z_{j i_j}^\top)^\top$.
Note that this assumption is stronger than the assumption of \textit{independent right-censoring} (\citealt{anbogikei93}, typically used in Cox regression), but it is commonly required for mathematical analyses of pseudo-observation-based regression techniques; cf.\ \cite{grageschu09} and \cite{ovpape17}.
It can be relaxed to covariate-dependent right-censoring if the employed nonparametric estimator admits an inverse-probability-of-censoring weighting structure and if the covariate's support is finite; see \cite{ovpape19}.

We define the Kaplan-Meier estimator for group $j$'s survival function $S_j, j=1,2,$ by
$$\hat S_j(t) = \prod_{i_j: X_{j(i_j)} \leq t}\Bigg(1 -  \frac{\sum_{\ell_j=1 }^{n_j} \delta_{j \ell_j} 1\{X_{j \ell_j} = X_{j (i_j)}\} }{\sum_{\ell_j=1}^{n_j}  1 \{ X_{j \ell_j} \geq X_{j (i_j)} \} } \Bigg),$$
where $X_{j(1)} < \dots < X_{j(m_j)}$, $m_j \le n_j$, denote the ordered distinct uncensored event times.

\begin{remark}
\label{rem:finite_tau}
In practice, longitudinal studies typically end after a finite time horizon, say $\tau >0$, before all events could have occurred.
This renders the right tail of the survival functions unidentifiable.
As a solution, there are mathematical approaches that allow the survival and censoring distributions to have the same (finite or infinite) support under stricter assumptions on the censoring distribution \citep{gill83,ying89,dobler19}.

One solution for the presently considered estimand $\theta$ is to instead use $\hat \theta(\tau) = - \int_0^\tau \hat S_1(u) d \hat S_2(u) $, which is a consistent estimator for $\theta(\tau) = P(\min(T_{1i_1},\tau) > \min(T_{2i_2}, \tau))$
if $P(T_{2i_2}=\tau)=0$. We will consider the estimand $\theta(\tau)$ and its estimator $\hat \theta(\tau)$ in the remainder of this paper. This is analogous to the Cox model where a time horizon $\tau$ should be chosen before computing the maximizer of the partial likelihood.
\end{remark}

The estimators, their jackknife variants, and pseudo-observations $\tilde \theta_{i_1 i_2}$ can be defined in just the same way as in the uncensored case (Eq.\@ \eqref{eq:pv}).
Also, in case of the identity link function $\mu(x)=x$, $\hat{\bs \beta}$ still exhibits the same structure as in~\eqref{eq:beta_hat_reduced}, except that the first entry in the vector, $\hat \theta$, has to be replaced by 
$$\tilde \theta_{\bullet \bullet} := \frac1{n_1n_2} \sum_{i_1=1}^{n_1} \sum_{i_2=1}^{n_2} \tilde \theta_{i_1 i_2}.$$

Let us now consider the GEE for defining the estimator $\hat{\bs \beta}$; the equation takes exactly the same form as~\eqref{eq:gee} and is therefore not restated here, but it is now based on the more general pseudo-observations resulting from right-censored data sets.

We are going to analyze the large-sample properties of $\hat{\bs \beta}$ as a consequence of some other convergences involving the pseudo-observations.
Before stating the central assumption and subsequently the main result (Assumption~\ref{ass:general} and Theorem~\ref{thm:censored} below), we introduce the notation $\bs Z_{i_1i_2} := (1,\bs Z_{1i_1}^\top, \bs Z_{2i_2}^\top)^\top$, which is a $p$-dimensional random vector.
All convergences are to be understood as $\min(n_1,n_2) \to \infty$.
\begin{assumption}
\label{ass:general}
\begin{enumerate}[leftmargin=1cm]
\item[] \phantom{a}\\[-.3cm]
    \item[(1.1) \ ] The function $\mu: \mathbb R \to \mathbb R$ is strictly increasing and it is three times continuously differentiable; \vspace{.08cm}
    \item[(1.2) \ ]
 for each $\bs \beta \in \mathbb R^p$ there exists the limit in probability of
 $$ C(\bs \beta) := \frac1{n_1n_2}\sum_{i_1=1}^{n_1} \sum_{i_2=1}^{n_2} (\tilde \theta_{i_1i_2} - \tfrac12 \mu_{i_1 i_2}) \,\mu_{i_1 i_2}, $$
 denoted by $f(\bs \beta)$, with $f: \mathbb R^p \to \mathbb R$ being twice continuously differentiable; \vspace{.08cm}
 \item[(1.3) \ ] for each $\bs \beta \in \mathbb R^p$ there exists the limit in probability of
 $$ \bs U(\bs \beta) = (\nabla C(\bs \beta))^\top = \frac1{n_1n_2}\sum_{i_1=1}^{n_1} \sum_{i_2=1}^{n_2} \bs A_{i_1i_2} (\tilde \theta_{i_1i_2} - \mu_{i_1 i_2}),$$
 denoted by $(\nabla f)^\top$,
 and a unique root $\bs \beta_0$ of $\nabla f$ satisfying $\nabla f(\bs \beta_0) = \bs 0$; \vspace{.08cm}
 \item[(1.4) \ ] there exists a neighborhood $B$ of $\bs \beta_0$ such that, for each $\bs \beta \in B$, the Jacobian
 $$ D \bs U(\bs \beta) = \frac1{n_1n_2}\sum_{i_1=1}^{n_1} \sum_{i_2=1}^{n_2} \Big(\bs Z_{i_1i_2} \bs Z_{i_1i_2}^\top \mu''(\bs \beta^\top \bs Z_{i_1i_2}) (\tilde \theta_{i_1i_2} - \mu_{i_1 i_2} ) - \bs A_{i_1i_2} \bs A_{i_1i_2}^\top \Big)$$
 converges in probability to some $(p \times p)$-matrix $- \check{\bs \Sigma}(\bs \beta)$, 
 which is negative definite in $\bs \beta_0$ and continuous in $\bs \beta_0$; further assume that the convergence in probability is uniform over $B$; \vspace{.08cm}
 \item[(1.5) \ ] $\sqrt{\frac{n_1n_2}{n_1+n_2}} \, \bs U(\bs \beta_0) \stackrel d \to \bs W \sim \mathcal{N}_p(\bs 0,  \check{\bs \Omega})$ for some $(p\times p)$-matrix $ \check{\bs \Omega}$.
 \end{enumerate}
\end{assumption}

\begin{theorem}
\label{thm:censored}
 Let $n_1, n_2 \to \infty$ such that $\frac{n_1}{n_1+n_2} \to \lambda \in (0,1)$.
 Under Assumption~\ref{ass:general},  the solution $\hat {\bs \beta}$ of \eqref{eq:gee}
\begin{itemize}[leftmargin = 1cm]
\item[(i) \ ] exists with a probability tending to 1, \vspace{.08cm}
\item[(ii) \ ] converges to $\bs \beta_0$ in probability, and \vspace{.08cm}
\item[(iii) \ ] satisfies
 $$ \sqrt{\frac{n_1 n_2}{n_1 + n_2}} \, (\hat {\bs \beta} - \bs \beta_0) \, \stackrel d {\longrightarrow}  \, \check{\bs \Sigma}(\bs \beta_0) \bs W \, \sim \,  \mathcal N_p(\bs 0, (\check{\bs \Sigma}(\bs \beta_0))^{-1} \check{\bs \Omega} ((\check{\bs \Sigma}(\bs \beta_0))^{-1})^\top). $$
\end{itemize}
\end{theorem}
A formal proof of Theorem~\ref{thm:censored} is given in Appendix \ref{app:proofs}.

\begin{remark}
    In the special case of fully observable data, $\tilde \theta_{i_1i_2} = 1\{T_{1i_1} > T_{2i_2}\}$, it is easy to verify Assumptions~\ref{ass:general}.2--\ref{ass:general}.5. 
    To see this, let us assume that $\mu$ is three times continuously differentiable and that the second moments of $\bs A_{i_1i_2}$ and $\bs A_{i_1i_2}(\tilde \theta_{i_1i_2} - \mu_{i_1 i_2})$ exist for all $\bs \beta \in \mathbb R^p$.
    The postulated assumptions are then consequences of laws of large numbers and a central limit theorem for two-sample $U$-statistics.
    
    For example, after noticing that $\text{E}(\tilde \theta_{i_1i_2} | \bs Z_{i_1}, \bs Z_{i_2}) = \mu_{i_1i_2}$,
    the Jacobian $D \bs U(\bs \beta_0) $ reduces to 
    $$-\frac1{n_1n_2}\sum_{i_1=1}^{n_1} \sum_{i_2=1}^{n_2} \bs A_{i_1i_2} \bs A_{i_1i_2}^\top + o_p(1) ,$$ 
    which converges in probability to $- \bs \Sigma(\bs \beta_0) = - \text{E}(\bs A_{i_1i_2} \bs A_{i_1i_2}^\top)$ by the strong law of large numbers for multi-sample $U$-statistics, e.g., Theorem~3.2.1 in \cite{kobo94}.
    It remains to assume that $- \bs \Sigma(\bs \beta_0)$ does not have a zero eigenvalue. Similarly, 
    the local uniformity of the convergence in Assumption~\ref{ass:general}.4 follows from the continuous differentiability of $D\bs U$ in~$\bs \beta_0$ and the pointwise convergence at each $\bs\beta$.
The existence of the stationary point $\bs \beta_0$ of the limiting function $f$ is ensured by the unbiasedness of~$\bs U$ at the true $\bs \beta = \bs \beta_0$ and the law of large numbers, from which it follows that $\bs U(\bs \beta_0) \stackrel{a.s.} {\longrightarrow} \bs 0 = \text{E}(\bs U(\bs \beta_0)) = \nabla f (\bs \beta_0) $. Finally, the convergence in distribution in Assumption~\ref{ass:general}.5  follows from a central limit theorem for $U$-statistics; cf.\@ Theorem~4.5.1 in \cite{kobo94}.
\end{remark}

\begin{remark}
    Based on Assumption~\ref{ass:general} and the consistency of $\hat{\bs \beta}$ for $\bs \beta_0$, a consistent estimator for $-\check{\bs \Sigma}(\bs \beta_0)$ is given by
    $$ \widehat{D \bs U}(\hat{\bs \beta}) := \frac1{n_1n_2} \sum_{i_1=1}^{n_1} \sum_{i_2=1}^{n_2} \Big(\bs Z_{i_1i_2} \bs Z_{i_1i_2}^\top \, \mu''(\hat{\bs \beta}^\top \bs Z_{i_1i_2}) \,(\tilde \theta_{i_1i_2} - \hat{\mu}_{i_1 i_2} ) - \hat{\bs A}_{i_1i_2} \hat{\bs A}_{i_1i_2}^\top \Big),$$
    where $\hat{\bs A}_{i_1i_2}$ and $\hat{\mu}_{i_1 i_2} $ are defined in the same way as ${\bs A}_{i_1i_2}$ and ${\mu}_{i_1 i_2} $, respectively, with $\bs \beta$ replaced by $\hat{\bs \beta}$.
    As in \cite{ovpape17}, we expect the limiting distribution to depend on a second derivative of the functional which produces the estimators $\hat \theta$ based on the nonparametric estimators $\hat S_1, \hat S_2$.
For this reason, we introduced another notation for $\check{\bs \Omega}$, in order to distinguish it from~$\bs \Omega$ in the uncensored case. In contrast to $\check{\bs \Omega}$, $\bs \Omega$ does not depend on such second derivative.
For practical applications, we propose to use a group-wise bootstrap procedure, {\color{black}i.e., randomly drawing with replacement from the original data points within each group,} instead of implementing an estimator of $\check{\bs \Omega}$. We will pursue this approach in the simulation study in Section~\ref{sec:simus} below.
In the uncensored case, it is fairly straightforward to estimate $\bs \Omega$ by an appropriate empirical estimator, as explained in Appendix \ref{app:proofs}.
\end{remark}

\section{Simulation study}
\label{sec:simus}

\vspace*{.1cm}
\subsection{Description of the simulation settings}
We conducted an extensive simulation study to evaluate the small-sample properties of the proposed methods.
Our main focus was on the small-sample values of the estimator $\hat {\bs \beta} =(\hat{\beta}_0, \hat{\beta}_{11} , \ldots , \hat{\beta}_{1p_1} , \hat{\beta}_{21} , \ldots , \hat{\beta}_{2p_2})^\top$ and the power of the hypothesis tests for ``$H_0^{(1)}: \beta_{11}=0$ vs.\@ $H_1^{(1)}: \beta_{11}\neq0$'' and ``$H_0^{(2)}: \beta_{21}=0$ vs.\@ $H_1^{(2)}: \beta_{21}\neq0$''. The latter tests (conducted at the 5\% significance level) investigated the influence of the first covariate on the Mann-Whitney effect. In all settings of the simulation study, we used the identity link function $\mu (x)=x$ and considered the unrestricted Mann-Whitney effect, setting $\tau = \infty$.

\begin{figure}[ht]
    \centering
    \includegraphics[width=0.4\linewidth]{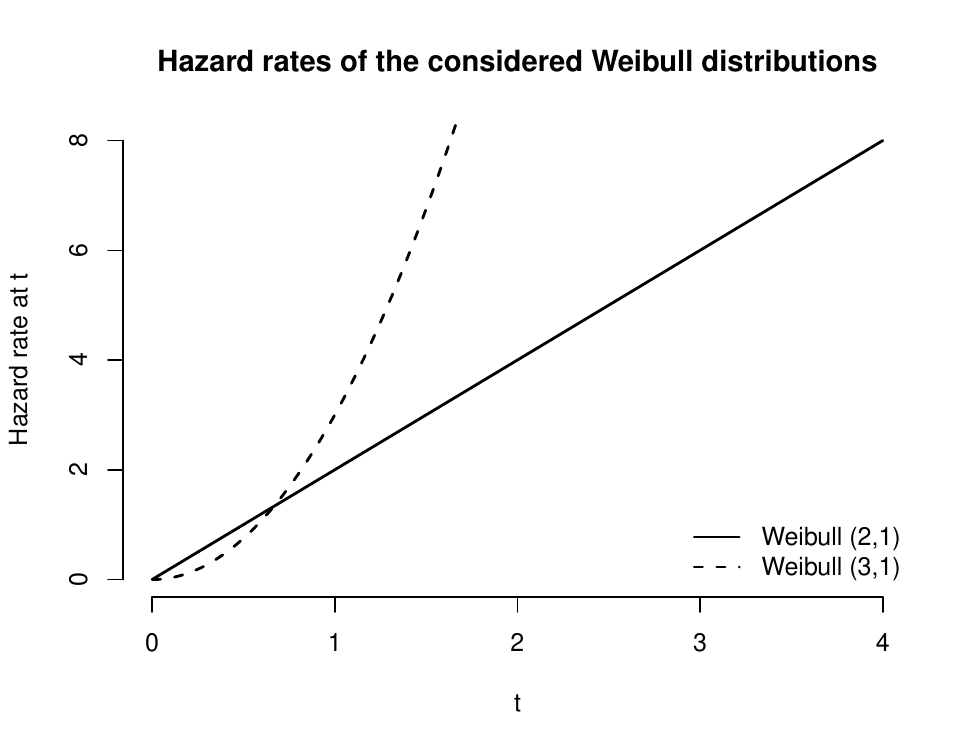}
    \includegraphics[width=0.4\linewidth]{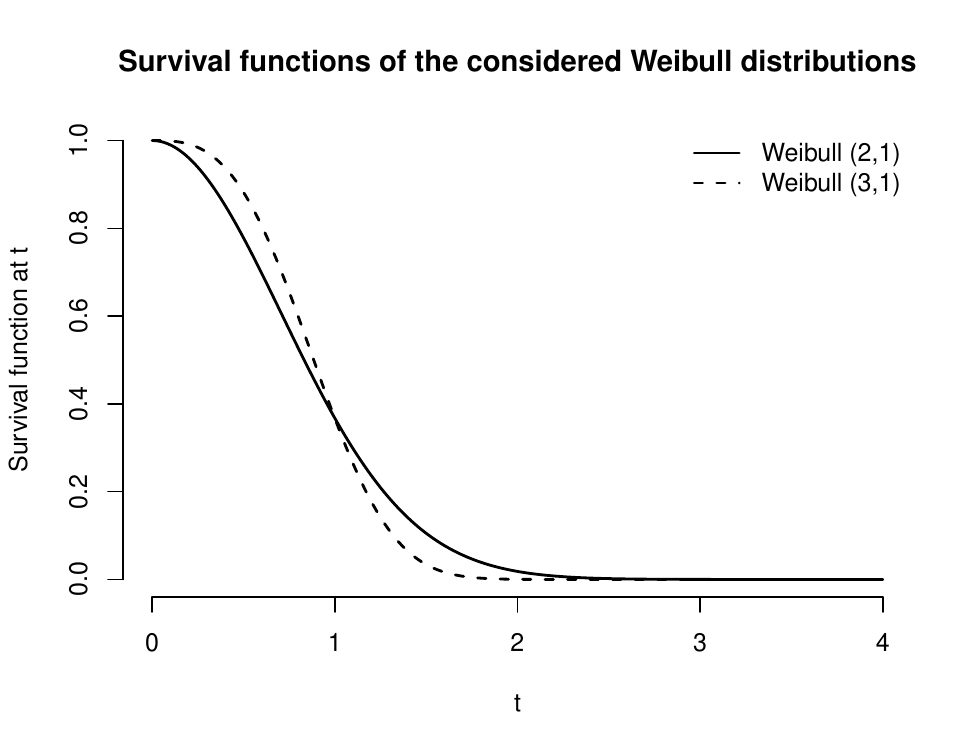}
    \vspace{-6mm}
    \caption{Hazard rates and survival functions of the Weibull$(2,1)$ and Weibull$(3,1)$ distributions. At the crossing point of the hazard rates ($t = 2/3 \cdot \lambda_{2i}^3 \,\lambda_{1i}^{-2}$ $= 2/3 \cdot 1^3 \cdot 1^{-2}$ $= 0.667$), the Weibull survival functions with shape parameters 2 and 3 assume the values $0.641$ and $0.744$, respectively.}
    \label{fig:hazards}
\end{figure}

The data-generating processes for the event and censoring times were defined as follows: For the 
event times, we specified Weibull distributions with survival functions $S_j(t|\bs Z_{ji}) = \exp(-(t/\lambda_{ji})^{k_j})$, $j=1,2$, with shape parameters $k_j > 0$ and scale parameters $\lambda_{ji} = \exp(\gamma_{j0} + \bs \gamma_j^\top \bs Z_{ji}) $.
The true parameter values $\gamma_0, \bs \gamma_1, \bs \gamma_2$ are specified in Table~\ref{tab:gamma_parms}. We address the relation between $\bs \beta_k$ and $\bs \gamma_k$, $k=0,1,2$, and in which cases the model is correctly specified, in Appendix~\ref{app:simu}. For the shape parameters, we considered two specifications: In {\em Setting I}, we specified $k_1=2$ and $k_2=3$, resulting in hazard rates crossing at $t=2/3 \cdot \lambda_{2i}^3 \,\lambda_{1i}^{-2}$. As an example, Figure~\ref{fig:hazards} shows the hazard rates and the survival functions of the considered Weibull  distributions for equal scale parameters $\lambda_1= \lambda_2=1$. In {\em Setting II}, we specified $k_1=k_2=3$, resulting in a Cox proportional hazards model. The distributions of the covariates (of dimensions $p_1=p_2 \in \{2,4\}$) were defined by a set of normal and binary distributions, as specified in 
Table~\ref{tab:cov_parms}. Censoring times were sampled independently using exponential distributions whose rate parameters  were adjusted such that the censoring proportion was either $25\%$, $50\%$, or $75\%$.
The values of the rate parameters were found with the help of an iterative procedure that either increased or decreased the rate depending on whether the empirical censoring proportion was below or above the targeted value.
All empirical proportions were derived from large data sets consisting of 1,000,000 independently sampled event and censoring times.
We also considered the fully observable case without censoring.

The sample sizes $(n_1, n_2)$ were set to $(200,200)$, $(150, 300)$, and $(500, 300)$. Ten thousand Monte Carlo replications were used for each combination of scale and shape parameter values, covariate numbers, censoring proportions, and sample sizes.

\begin{table}[ht]
\begin{center}
\caption{True parameter values $\gamma_{j0}, \bs \gamma_j$, $j=1,2$, used in the simulation study. Note that the regression model for the pseudo-observation approach is misspecified in Scenarios ii and iv. In Appendix~\ref{app:simu} we provide details on the characteristics of the proposed model for Weibull-distributed event times with $k_1=k_2$.
\label{tab:gamma_parms}} \vspace*{.1cm}
 \begin{tabular}{ccccccll}
 \hline
  Scenario & $(p_1,p_2)$ & $\gamma_{10}$ & $\gamma_{20}$ & $\bs \gamma_1^\top$ & $\bs \gamma_2^\top$ & True hypotheses & Model specification \\[1mm] \hline
  i & $(2,2)$ & $0$ & $0$ & $(0,0)$ & $(0,0)$ & $H_0^{(1)}$ and $H_0^{(2)}$ & correctly specified \\[1mm] 
  ii & $(2,2)$ & $0$ & $0$ & $(0.2,0)$ & $(0,0.5)$ & $H_1^{(1)}$ and $H_0^{(2)}$ & misspecified \\[1mm] 
  iii & $(4,4)$ & $0$ & $0$ & $(0,0,0,0)$ & $(0,0,0,0)$ & $H_0^{(1)}$ and $H_0^{(2)}$ & correctly specified \\[1mm] 
  iv & $(4,4)$ & $0 $& $0$ & $(0,0.2,0.4,0.6)$ & $(-0.2,0.4,-0.6,0)$ & $H_0^{(1)}$ and $H_1^{(2)}$ & misspecified \\[1mm] \hline
\end{tabular}
\end{center}
\end{table}

\begin{table}[ht]
\begin{center}
\caption{Distributions of the covariates $(Z_{j1},Z_{j2})$ for $p_j=2$ and $(Z_{j1}, \dots, Z_{j4})$ for $p_j=4$ in groups $j=1,2$. \label{tab:cov_parms}} \vspace*{.1cm}
 \begin{tabular}{ccc}
 \hline
  $j$ & $p_j$ & Distribution \\[1mm] \hline
  1 & 2 & $Z_{11} \sim \mathcal{N}_1(0,1)$, \ $Z_{12} \sim \text{Bin}(1,0.5 + 0.1 \cdot \text{sign}(Z_{11}))$ \\[1mm] 
  2 & 2 & $Z_{21} \sim \mathcal{N}_1(0,1.2)$, \ $Z_{22} \sim \text{Bin}(1,0.7 - 0.05 \cdot \text{sign}(Z_{21}))$  \\[1mm] 
  1 & 4 & $(Z_{11},Z_{12}) \sim \mathcal{N}_2 \big( (0,0)^\top, \big(\begin{smallmatrix} 1.0 & 0.2 & \\ 0.2 & 1.0\end{smallmatrix}\!\! \big)\big)$, \ $Z_{13} \sim \text{Bin}(1, 0.4)$, \ $Z_{14}\sim \text{Bin}(1, 0.6)$ \\[1mm] 
  2 & 4 & $(Z_{21},Z_{22}) \sim \mathcal{N}_2 \big( (0,0)^\top, \big(\begin{smallmatrix} 1.1 & 0.3 & \\ 0.3 & 1.1\end{smallmatrix}\!\! \big)\big)$, \ $Z_{23}, Z_{24} \sim \text{Bin}(1, 0.5 + 0.1 \cdot \text{sign}(Z_{21}))$ \\[1mm] 
\hline
\end{tabular}
\end{center}
\end{table}

\subsection{Description of the hypothesis tests}
\label{sec:descTests}

The tests for $H_0^{(j)}: \beta_{j1}=0$ vs.\ $H_1^{(j)} : \beta_{j1}\neq 0$  were conducted in four different ways. The first test used the empirical standard deviation of the bootstrapped parameter estimates, denoted by $\hat\sigma^*_{j,\text{emp}}$: $$ \varphi_{j,\text{emp}} := 1\{|\hat \beta_{j1}| / \hat\sigma^*_{j,\text{emp}} > z_{0.975} \}, \quad j=1,2, $$
where $z_{0.975}$ denotes the $97.5\%$-quantile of the standard normal distribution. The second test used the inter-quartile range (IQR) of the bootstrapped parameter estimates $\hat \beta_{j1}^{(b)*}, b=1,\dots, B$, divided by 1.349:
$$ \varphi_{j,\text{IQR}} := 1\{|\hat \beta_{j1}| / \hat\sigma^*_{j,\text{IQR}} > z_{0.975} \}, \quad j=1,2, $$
where $\hat\sigma^*_{j,\text{IQR}} = |\hat \beta_{j1,(75\%)}^* - \hat \beta_{j1,(25\%)}^*|/1.349$. By definition, $\hat\sigma^*_{j,\text{IQR}} $ corresponds to an alternative estimator of the standard deviation for normally distributed data.
Analogously, the third test used the median absolute deviation (MAD) of the bootstrapped parameter estimates, multiplied by 1.483:
$$ \varphi_{j,\text{MAD}} := 1\{|\hat \beta_{j1}| / \hat\sigma^*_{j,\text{MAD}} > z_{0.975} \}, \quad j=1,2, $$
where $\hat\sigma^*_{j,\text{MAD}} = \textnormal{med}( |\hat \beta_{j1}^{(\tilde b)*} - \textnormal{med}(\hat \beta_{j1}^{(b)*}: b=1,\dots, B )| :  \tilde b=1,\dots, B) \cdot 1.483$. The fourth test was based on the percentile method, i.e., it rejects the null hypothesis if $\hat \beta_{j1}$ is contained in either of the 2.5\% tails of the distribution of the centered bootstrap parameter estimates. This leads to the test
$$ \varphi_{j,\text{quantile}} = 1\{ \hat \beta_{j1} < q^*_{j,2.5\%} \} + 1\{ \hat \beta_{j1} > q^*_{j,97.5\%} \}, \quad j=1,2, $$
where $q^*_{j,2.5\%} = (\hat \beta_{j1}^{(b)*} - \hat \beta_{j1})_{(2.5\%)} $ and $q^*_{j,97.5\%} = (\hat \beta_{j1}^{(b)*} - \hat \beta_{j1})_{(97.5\%)} $.

We used the {\em warp-speed bootstrap} \citep{gipowhi13} to conduct the tests described above. The warp-speed bootstrap is particularly useful for accelerating extensive simulation studies; in every Monte Carlo simulation run it draws only one bootstrap sample based on which the estimators are re-calculated.
In the final step, the collection of all bootstrapped parameter estimates, say $\hat{\bs \beta}_{k}^{(b)*}$, $k=0,1,2$, $b=1,\ldots,$10,000, is used for the inferential procedures.

\subsection{Comparison to the Cox proportional hazards model}

In a further step, we compared the performance of the proposed method to $z$-tests obtained from Cox regression. To this end, we fitted a Cox model to each of the Monte Carlo samples, using the group indicator as an additional covariate. Also, to obtain a model serving a similar purpose as the model for the Mann-Whitney effect, we included all $p_1=p_2$ interactions between the group indicator and the other $p_1=p_2$ covariates. It can be shown that testing for $H_0^{(j)}: \beta_{j1}=0$ vs.\ $H_1^{(j)} : \beta_{j1}\neq 0$, $j=1,2$, in the pseudo-observation-based model translates to testing for specific linear combinations in the parameters of the Cox model. Details on the derivation and interpretation of the Mann-Whitney effect in a Cox regression model are provided in Appendix \ref{app:Cox}.

\subsection{Results of the simulation study}

\begin{figure}[b!]
    \centering
    \includegraphics[width = 0.9\linewidth]{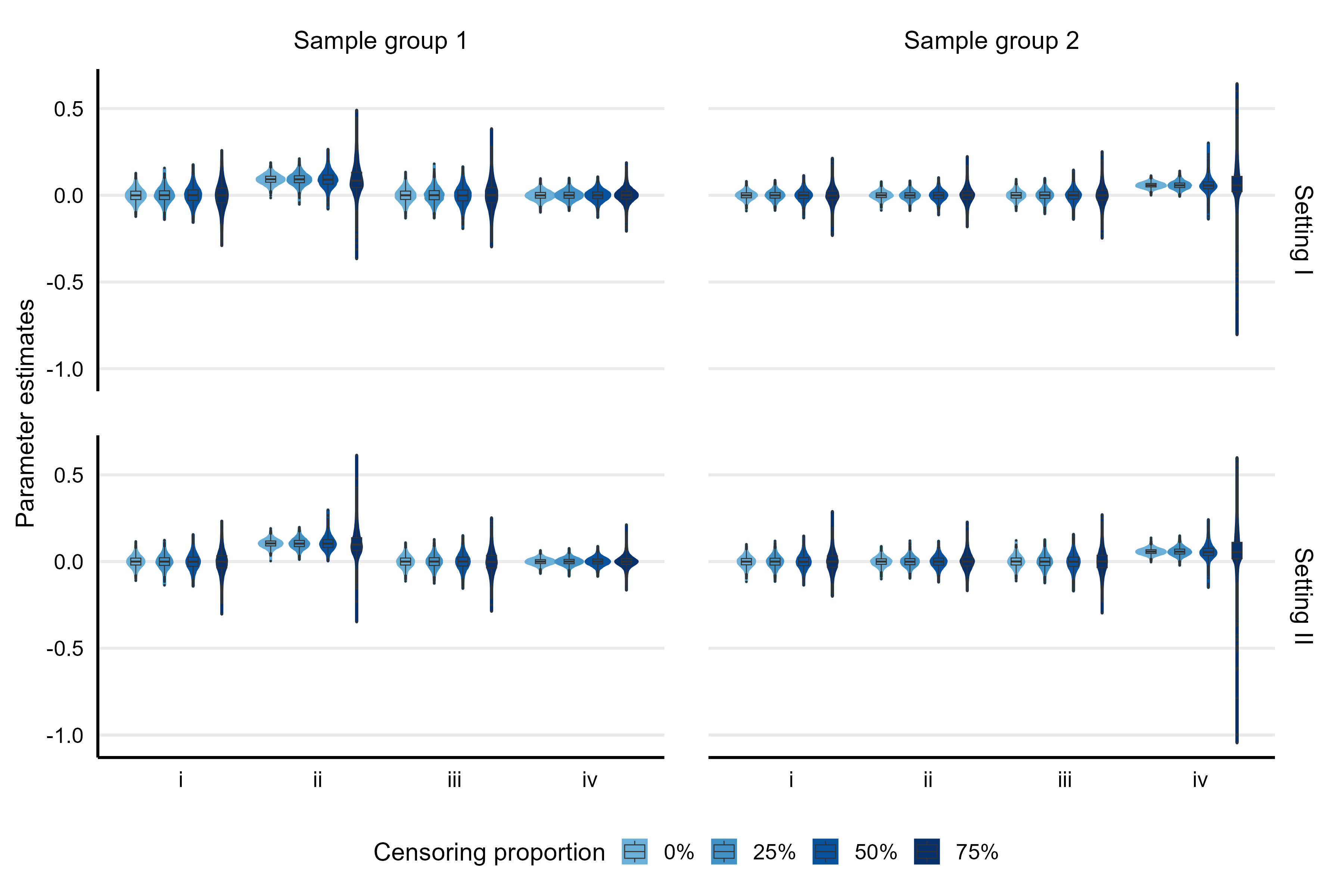}
    \caption{Results from the simulation study for sample sizes $n_1=n_2=200$. The figure presents boxplots of the estimates of $\beta_{j1}$, $j=1,2$, for different covariate space dimensions $p\in \{2,4\}$. Settings I and II refer to unequal and equal  Weibull shape parameters, respectively.}
    \label{fig:boxplots_betas}
\end{figure}

\begin{figure}[h!]
    \centering
    \includegraphics[scale = 0.55]{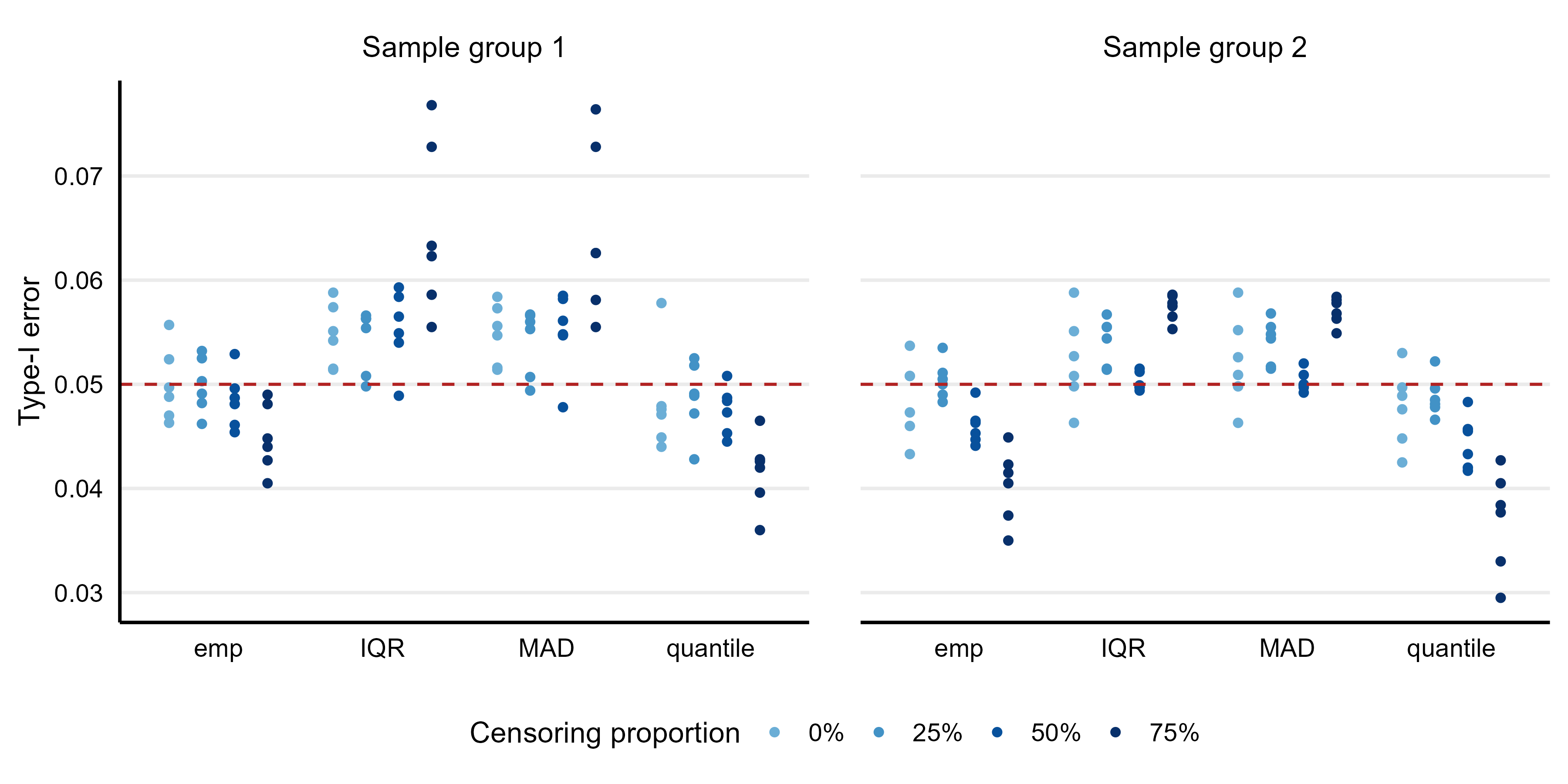}
    \caption{Type-I error rates of the tests for  $H_0^{(j)}:\beta_{j1}=0$ vs.\ $H_a^{(j)}: \beta_{j1}\neq 0$, $j=1,2$,  across all settings for sample sizes $n_1=n_2=200$. The nominal significance level $\alpha=5\%$ is displayed as a horizontal dashed line. The labels on the x-axes correspond to the four tests $\varphi_{j,\text{emp}}$, $\varphi_{j,\text{IQR}}$, $\varphi_{j,\text{MAD}}$ and $\varphi_{j,\text{quantile}}$.}
\label{fig:boxplots_type-I_power}
\end{figure}

Figure~\ref{fig:boxplots_betas} displays boxplots of the estimated parameter values of $\beta_{11}$ and $\beta_{21}$ for sample sizes $n_1=n_2=200$ and for the cases $p_1=p_2=2$ (Scenarios i and ii in Table \ref{tab:gamma_parms}) and $p_1=p_2=4$ (Scenarios iii and iv in Table \ref{tab:gamma_parms}). The plots in Figure~\ref{fig:boxplots_type-I_power} illustrate the corresponding estimated type-I error rates of the novel tests.  Table~\ref{tab:rejection} contains all rejection rates obtained from the simulation study for $n_1=n_2=200$,
both under the null hypotheses and the alternative hypotheses.
The corresponding results for $(n_1,n_2) \in \{ (150,300), (500,300)\}$ can be found in Appendix~\ref{app:additional_simulation}.

\begin{table}[ht]
\begin{center}
\caption{Results from the simulation study for sample sizes $n_1=n_2=200$. The table presents the estimated rejection rates (rounded to full percentage points) of the tests for $H_0^{(j)}:\beta_{j1}=0$ vs.\ $H_1^{(j)}: \beta_{j1}\neq 0$, $j=1,2$, and the corresponding $z$-tests in the Cox model for the presence of a group-covariate interaction (Scenarios i and ii: $p_1=p_2=2$; Scenarios iii and iv: $p_1=p_2=4$).
The underlined numbers indicate estimated powers under $H_1^{(j)}$.
Numbers are printed in bold when contained in the 95\%-Monte-Carlo error interval $[4.58\%,5.43\%]$ under $H_0^{(j)}$ or when representing the highest power under~$H_1^{(j)}$.
The abbreviations emp, IQR, MAD and quantile refer to the four tests $\varphi_{j,\text{emp}}$, $\varphi_{j,\text{IQR}}$, $\varphi_{j,\text{MAD}}$ and $\varphi_{j,\text{quantile}}$, respectively. \label{tab:rejection}}
\vspace{2mm}
\scriptsize
\begin{tabular}{cccc|rrrr|r|rrrr|r}
  \hline
 &Weibull & $(\boldsymbol \gamma_1, \boldsymbol \gamma_2)$ & censoring & \multicolumn{5}{c|}{$H_0^{(1)}$ vs.\ $H_1^{(1)}$} & \multicolumn{5}{c}{$H_0^{(2)}$ vs.\ $H_1^{(2)}$} \\ 
 $(p_1,p_2)$ & shapes $(k_1,k_2)$ & scenario & proportion & emp & IQR & MAD & quantile & Cox & emp & IQR & MAD & quantile & Cox \\
  \hline
(2,2) & (2,3) & i & 0 & 6 & 6 & 6 & 6 & 8 & \textbf{5} & \textbf{5} & \textbf{5} & \textbf{5} & 9 \\
   &(3,3)& i & 0 & \textbf{5} & 6 & 6 & \textbf{5} & 5 & \textbf{5} & \textbf{5} & \textbf{5} & 4 &  5 \\
   &(2,3)& ii & 0 & \underline{100} & \underline{100}& \underline{100} & \underline{100} & \underline{72} & \textbf{5} & 6 & 6 & \textbf{5} & {11} \\
   &(3,3)& ii & 0 & \underline{100} & \underline{100} & \underline{100} & \underline{100} & \underline{90} & \textbf{5} & 6 & 6 & \textbf{5} & {5} \\
   &(2,3)& i & 25 & \textbf{5} & \textbf{5} & \textbf{5} & \textbf{5} & 8 & \textbf{5} & {5} & {5} & \textbf{5} & 9 \\
   &(3,3)& i & 25 & \textbf{5} & 6 & 6 & \textbf{5} & \textbf 5 & \textbf{5} & 6 & 6 & \textbf{5} & 6 \\
   &(2,3)& ii & 25 & \underline{99} & \underline{99} & \underline{99} & \underline{99} & \underline{59} & \textbf{5} & 6 & 6 & \textbf{5} & {11} \\
   &(3,3)& ii & 25 & \underline{100} & \underline{100} & \underline{100} & \underline{100} & \underline{81} & \textbf{5} & \textbf{5} & \textbf{5} & \textbf{5} & {5} \\
   &(2,3)& i & 50 & \textbf{5} & \textbf{5} & {5} & \textbf{5} & 8 & 4 & \textbf{5} & \textbf{5} & 4 & 10 \\
   &(3,3)& i & 50 & \textbf{5} & 6 & 6 & \textbf{5} & 6 & \textbf{5} & \textbf{5} & \textbf{5} & {5} & 5 \\
   &(2,3)& ii & 50 & \underline{91} & \underline{93} & \underline{93} & \underline{91} & \underline{44} & \textbf{5} & \textbf{5} & \textbf{5} & {5} & {10} \\
   &(3,3)& ii & 50 & \underline{99} & \underline{100} & \underline{100} & \underline{99} & \underline{63} & {5} & \textbf{5} & \textbf{5} & 4 & {5} \\ 
   &(2,3)& i & 75 & 4 & 6 & 6 & 4 & 8 & 4 & 6 & 6 & 3 & 8 \\
   &(3,3)& i & 75 & 4 & 6 & 6 & 4 & \textbf 5 &  4 & 6 & 6 & 4 & 5 \\
   &(2,3)& ii & 75 & \underline{30} & \underline{46} & \underline{46} & \underline{27} & \underline{25} & 4 & 6 & 6 & 3 & {9} \\
   &(3,3)& ii & 75 & \underline{54} & \underline{73} & \underline{73} & \underline{47} & \underline{38} & 4 & 6 & 6 & 4 & {6} \\ \hline \\[-.3cm]
(4,4) &(2,3)& iii & 0 & \textbf{5} & \textbf{5} & \textbf{5} & \textbf{5} & 7 & \textbf{5} & \textbf{5} & \textbf{5} & \textbf{5} & 9 \\
   &(3,3)& iii & 0 & \textbf{5} & \textbf{5} & \textbf{5} & 4 & \textbf 5 & 4 & \textbf{5} & \textbf{5} & 4 & 6 \\
   &(2,3)& iv & 0 & \textbf{5} & \textbf{5} & {5} & \textbf{5} & 8 & \underline{100} & \underline{100} & \underline{100} & \underline{100} & \underline{100} \\
   &(3,3)& iv & 0 & \textbf{5} & 6 & 6 & 4 & 5 & \underline{100} & \underline{100} & \underline{100} & \underline{100} & \underline{100} \\
   &(2,3)& iii & 25 & \textbf{5} & \textbf{5} & \textbf{5} & \textbf{5} & 7 & \textbf{5} & \textbf{5} & \textbf{5} & \textbf{5} & 9 \\
   &(3,3)& iii & 25 & \textbf{5} & 6 & 6 & \textbf{5} & 6 & \textbf{5} & 6 & {5} & \textbf{5} & 6 \\
   &(2,3)& iv & 25 & \textbf{5} & 6 & 6 & \textbf{5} & 8 & \underline{99} & \underline{99} & \underline{99} & \underline{99} & \underline{100} \\
   &(3,3)& iv & 25 & \textbf{5} & 6 & 6 & 4 & 5 & \underline{98} & \underline{98} & \underline{98} & \underline{98} & \underline{100} \\
   &(2,3)& iii & 50 & {5} & \textbf{5} & \textbf{5} & 4 & 7 & 4 & \textbf{5} & \textbf{5} & 4 & 10 \\
   &(3,3)& iii & 50 & \textbf{5} & {5} & {5} & \textbf{5} & 6 & \textbf{5} & \textbf{5} & \textbf{5} & \textbf{5} & 6 \\
   &(2,3)& iv & 50 & \textbf{5} & 6 & 6 & {5} & 8 & \underline{64} & \underline{79} & \underline{79} & \underline{56} & \underline{97} \\
   &(3,3)& iv & 50 & \textbf{5} & 6 & 6 & \textbf{5} & 6 & \underline{55} & \underline{71} & \underline{71} & \underline{51} & \underline{100} \\
   &(2,3)& iii & 75 & \textbf{5} & 6 & 6 & \textbf{5} & 7 & 4 & 6 & 6 & 4 & 9 \\
   &(3,3)& iii & 75 & 4 & 6 & 6 & 4 & 6 & 4 & 6 & {5} & 4 & 6 \\
   &(2,3)& iv & 75 & \textbf{5} & 7 & 7 & 4 & 8 & \underline{11} & \underline{27} & \underline{27} & \underline{8} & \underline{76} \\
   &(3,3)& iv & 75 & 4 & 8 & 8 & 4 & 6 & \underline{9} & \underline{26} & \underline{26} & \underline{7} & \underline{96} \\
   \hline
\end{tabular}
\end{center}
\end{table}

From the boxplots in Figure~\ref{fig:boxplots_betas} it is seen that the parameter estimates concentrate around 0 quite symmetrically in all cases when there is no covariate effect on the Mann-Whitney effect (cf.\@ Table~\ref{tab:gamma_parms}).
Whenever there is an effect, there is an obvious shift in the boxplot, as desired. 
The plots in Figure~\ref{fig:boxplots_type-I_power} suggest that the tests based on the empirical standard deviation and the empirical quantiles of the bootstrap estimators control the type-I error rate quite well; nearly all estimated rejection rates are between 3.5\% and 5\%. On the other hand, these two tests are slightly conservative, especially for higher censoring proportions. 
As we will see below, this is connected to a suboptimal power.

In contrast, the two tests based on IQR and MAD exhibit a too liberal behavior; most of the estimated rejection rates are between 5\% and 8\%, exceeding the nominal level $\alpha = 5\%$.
Both tests arrive at very similar decisions.
The IQR- and MAD-based tests seem particularly liberal under model misspecification (Scenario iv) combined with a high censoring proportion (75\%), see Table~\ref{tab:rejection}.


Table \ref{tab:rejection} also suggests that the considered sample sizes are likely too small to guarantee a satisfactory power of the  tests when the censoring proportion is high (75\%). Furthermore, the tests based on the empirical quantiles and the empirical standard deviation are clearly less powerful than the tests based on the IQR and the MAD. 
This could be partially explained by the conservative behavior of the first two and the liberal behavior of the latter two tests under the null hypothesis. Apart from the settings with a high censoring proportion (75\%), however, all tests show a satisfactory power. For this reason, and in view of the better control of the type-I error rate, the tests based on the empirical quantiles and the standard deviation still seem preferable. In our analysis of the SUCCESS-A study data (involving a much larger sample size, see Section \ref{sec:data}), we found almost no differences between the four tests.

In Table~\ref{tab:rejection} we also present the estimated type-I error rates and the estimated power of the $z$-test based on the Cox model.
Since this test is not based on a resampling procedure, it is unsurprising that the type-I error rates are generally not as close to 5\% as those of the four novel tests. In fact, the $z$-test based on the Cox model seems too liberal in many of the considered scenarios. On the other hand, the proposed tests based on the pseudo-observation approach are often less powerful than the $z$-test based on the Cox model.
In this respect, it should be noted that the proposed tests based on pseudo-observations focus on parameters that differ from those in the Cox model. As a consequence, a direct comparison of the tests is challenging.

In conclusion, the simulation study provides evidence that the proposed tests based on the two-sample pseudo-observation approach exhibit satisfactory power in small-sample settings, provided that the censoring proportion does not exceed~50\%. They also keep up with the $z$-test in the Cox model, which, however, failed to control the type-I error in many of the considered settings.
Among the novel tests, the two tests based on either a studentization by means of the empirical standard deviation or the empirical quantiles of the bootstrapped parameter estimates seem most reliable in terms of type-I error control -- albeit having slightly lower power compared to the other novel tests.

\section{Analysis of the SUCCESS-A study data}
\label{sec:data}

We illustrate our approach using data from the multicenter randomized phase III SUCCESS-A trial (NCT02181101), which enrolled 3,754 patients with primary invasive breast cancer between September 2005 and March 2007 \citep{GregorioHaeberleFaschingetal.2020}. Participants were randomized 1:1 to receive either standard chemotherapy plus gemcitabine (intervention group) or standard chemotherapy alone (control group). The primary endpoint was disease-free survival (DFS), as defined in Section~\ref{sec:intro}. Because death was included in the  definition of DFS, it was not treated as a competing event in this secondary analysis. For further details on the inclusion/exclusion criteria and the design of the study, we refer to \cite{GregorioHaeberleFaschingetal.2020}. 

Patients were censored at their last known disease-free date, resulting in an event proportion of 12.2\% (458 events among 3,754 patients). The maximum follow-up duration was 5.5 years (6 months of chemotherapy plus 5 years of follow-up), with a median follow-up of 5.2 years. Patient and tumor characteristics included age at randomization, body mass index (BMI), menopausal status, tumor stage, tumor grade, lymph node status, tumor type, HER2 status, estrogen receptor (ER) status, and progesterone receptor (PR) status. After excluding individuals with missing covariate data, the final dataset used for analysis comprised 3,652 patients.
Appendix~\ref{app:success} contains tables with more detailed data summaries and results from supplementary analyses.

In the original SUCCESS-A study, no overall treatment effect on disease-free survival was observed. However, exploratory analyses using a Cox regression model with treatment–covariate interaction terms suggested possible treatment effects in specific patient subgroups \citep{GregorioHaeberleFaschingetal.2020}. Hence, our objective was to explore the heterogeneity of treatment effects by identifying patient subgroups that may show a higher benefit from the intervention compared to control therapy, using a multivariable framework to account for the joint and potentially correlated effects of patient and tumor characteristics. To this end, we fitted a model of the form $P(\min(T_1, \tau) > \min(T_2, \tau) \mid \bs Z_1, \bs Z_2) = \mu(\beta_0 + \bs\beta_1^\top \bs Z_1 + \bs\beta_2^\top \bs Z_2)$ for DFS using the above described approach. Here, the subscripts $1$ and $2$ refer to the intervention and control groups, respectively. Pseudo-observations were computed as defined in Equation \eqref{eq:pv} and Remark~\ref{rem:finite_tau} using $\tau = 5.5$ years. 
Ninety-five percent confidence intervals were derived using the four methods described in Section \ref{sec:descTests} (based on $B=$ 2,000 bootstrap replications). Analogous to the simulation study in Section \ref{sec:simus}, we used the identity link function $\mu(x) = x$. Furthermore, we considered the special case $\bs Z_1 = \bs Z_2 = \bs Z$ (corresponding to equal patient and tumor characteristics in both groups). Hence, the model reduced to  $P(\min(T_1, \tau) > \min(T_2, \tau) \mid \bs Z) = \beta_0 + (\bs\beta_1 + \bs\beta_2)^\top \bs Z$. 

In our analysis, we were interested in estimating the probability $P(\min(T_1,\tau) > \min(T_2,\tau) \mid \bs Z)$. However, this probability could be very small and thus give a wrong impression of the treatment's effectiveness when there is a substantial number of tied event times and/or when many patients are censored at $\tau$. In the SUCCESS-A data, a large number of ties were observed at $\tau = 5.5$ years (940/3652 = 25.74\%). To account for these ties, we defined and applied a \emph{tie-corrected estimator} as follows: Let $\hat{S}_1(t)$ and $\hat{S}_2(t)$ denote the Kaplan-Meier survival estimates in the two groups, and $\Delta \hat{S}_1(t) = \hat{S}_1(t-) - \hat{S}_1(t)$ and $\Delta \hat{S}_2(t) = \hat{S}_2(t-) - \hat{S}_2(t)$ the jumps in $\hat{S}_1$ and $\hat{S}_2$, respectively, at time $t$. (Note that $\Delta \hat{S}_1(t)$ and $\Delta \hat{S}_2(t)$ are zero if there is no uncensored event time equaling $t$.) To correct for ties, we estimated $0.5 \cdot P(\min(T_1,\tau) = \min(T_2,\tau) \mid \bs Z = \bs z) + P(\min(T_1,\tau) > \min(T_2,\tau) \mid \bs Z = \bs z) $ by
\begin{align}
\label{eq:predictions}
    & 0.5 \cdot \hat{P}(\min(T_1,\tau) =  \min(T_2,\tau) \mid \bs Z = \bs z) + \hat{P}(\min(T_1,\tau) > \min(T_2,\tau) \mid \bs Z = \bs z) \nonumber\\
                       & = \ 0.5 \cdot \Big[ \hat{S}_1(\tau) \cdot \hat{S}_2(\tau) + \sum_{t \le \tau} \Delta \hat{S}_1(t) \cdot \Delta \hat{S}_2(t) \Big] + \hat{\beta}_0 + (\hat{\beta}_1 + \hat{\beta}_2)^\top \bs z . 
\end{align}
The rationale for this nonparametric correction of $\hat{\beta}_0 + (\hat{\beta}_1 + \hat{\beta}_2)^\top \bs z$ is as follows: The sum in (\ref{eq:predictions}) captures the probability of exact ties at discrete event times $t \le \tau$, while the product $\hat{S}_1(\tau)\cdot \hat{S}_2(\tau)$ accounts for ties arising from censoring at the end of follow-up. Since it is reasonable to assign half of the probability to $T_1 > T_2$ and the other half to $T_1 < T_2$ in the presence of a tie, we multiplied this term by $0.5$. Applying the tie-corrected estimator, the resulting correction term for the SUCCESS-A data was estimated as $0.3632$. 

Analogous to the simulation study, we compared the proposed method to a Cox proportional hazards model with group-by-covariate interaction effects (for details on the specification of this model, see Appendix~\ref{app:Cox}).

\begin{figure}
    \centering
    \includegraphics[scale = 0.5]{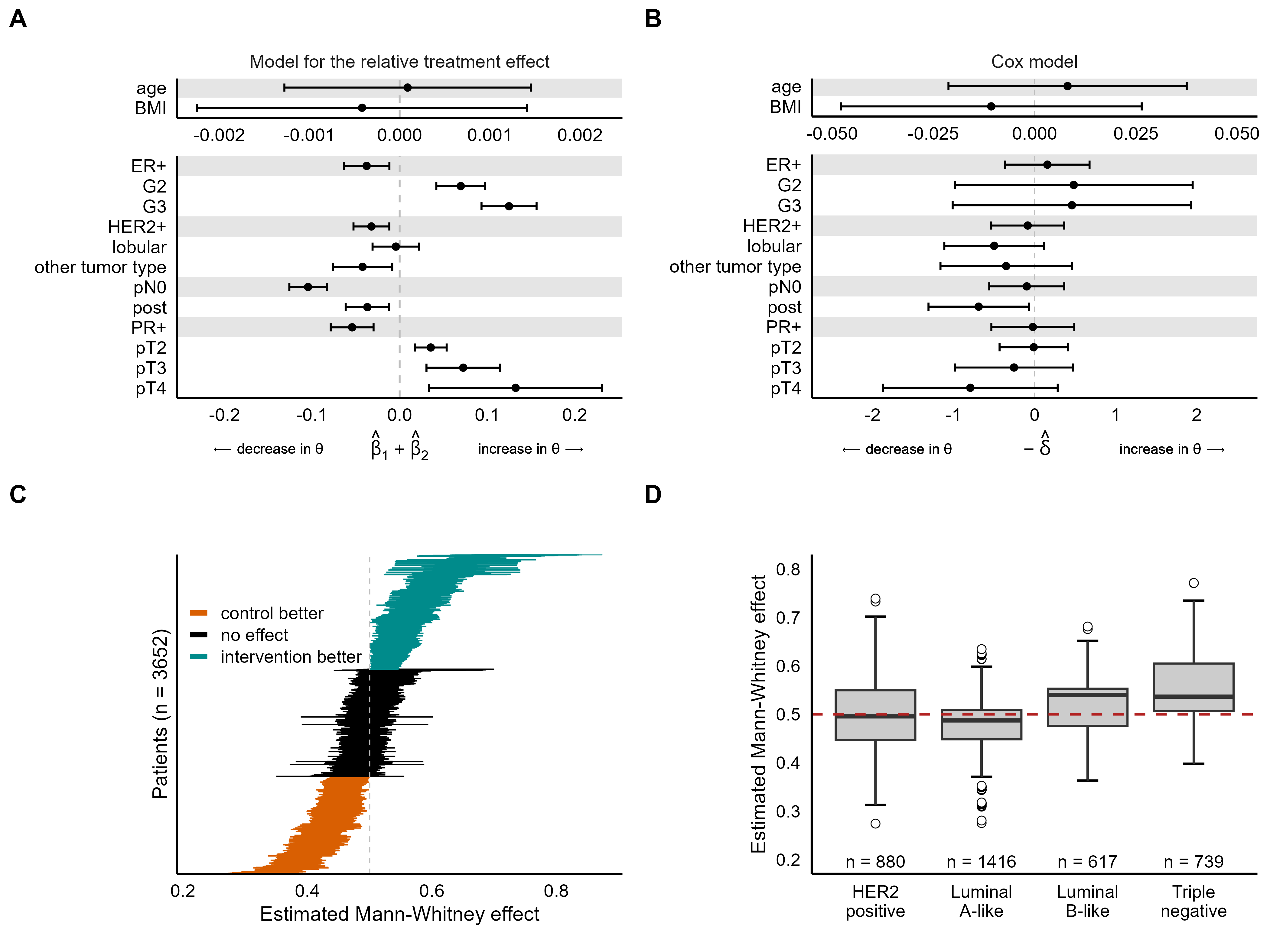}
    \caption{Analysis of the SUCCESS-A study data. \textbf{(A)} Estimated coefficients $\hat{\bs\beta}_1 + \hat{\bs\beta}_2 $ with 95\% confidence intervals, obtained using the empirical standard deviation method. {\color{black}The confidence intervals obtained from the IQR, MAD, and empirical quantile methods are presented in Figure \ref{fig:supp_coefficients}.} Positive values of $\hat{\bs\beta}_1 + \hat{\bs\beta}_2$ indicate a higher estimated {\color{black}probability of $\min(T_1, \tau) > \min(T_2, \tau)$,} corresponding to a higher expected benefit of the intervention compared to the reference categories of the covariates. Age and BMI are measured in years and~$kg/m^2$, respectively. The reference categories are ER$-$, G1, HER2$-$, ductal, pN+, pre-menopausal, PR$-$ and pT1. The estimated value of the intercept $\beta_0$ is 0.1560 [0.0797; 0.2322]. \textbf{(B)} {\color{black}Estimated coefficients of the group-by-covariate interactions $\delta_1, \ldots , \delta_{p_1}$ with corresponding 95\%~confidence intervals, as obtained from the Cox proportional hazards model. The control group serves as the reference group. Consequently, as demonstrated in Appendix \ref{app:Cox}, negative interaction effects indicate a higher expected benefit of the intervention compared to the reference categories of the covariates. For better comparability with Panel \textbf{(A)}, we multiplied all coefficient estimates by $-1$.} \textbf{(C)} Estimated tie-corrected truncated Mann-Whitney effects for all patients {\color{black}(as obtained from Eq.~\eqref{eq:predictions})}, grouped by estimated benefit in the intervention and control groups (using a probability threshold of 0.5). \textbf{(D)} Estimated tie-corrected truncated Mann-Whitney effects {\color{black}(as obtained from Eq.~\eqref{eq:predictions})} stratified by molecular tumor subtype, highlighting the heterogeneity of Mann-Whitney effects across subgroups.}
    \label{fig:dataexample}
\end{figure}

Figure \ref{fig:dataexample}A visualizes the estimated coefficients $\hat{\bs\beta}_1 + \hat{\bs\beta}_2 $  with 95\% confidence intervals, as obtained using the empirical standard deviation of the bootstrapped estimates. The combined coefficients represent the contributions of the covariates to the conditional probability of $\min(T_1, \tau) > \min(T_2, \tau)$. Accordingly, values of $\hat{\bs\beta}_1 + \hat{\bs\beta}_2 $ larger than $0$ indicate higher values of the Mann-Whitney effect compared to the reference categories of the covariates (as positive coefficients increase $\hat{P}(\min(T_1, \tau) > \min(T_2, \tau) \mid \bs Z)$). For example, our model suggests a higher benefit of the intervention in patients with G2 (0.0698 [0.0419; 0.9766]) and G3 (0.1249 [0.0934; 0.1563]) tumors compared to G1 tumors. Notably, this effect was not identified in the original publication. Furthermore, our model indicates a higher benefit of the control treatment among patients with non-affected lymph nodes (pN0) compared to patients with affected lymph nodes (pN+), with an estimated effect of $-$0.1045 [$-$0.1259; $-$0.0832]. In the original publication, the estimated effect pointed in the same direction but had a much wider confidence interval (HR = 1.11 [0.70; 1.74], cf.\@ \citealt{GregorioHaeberleFaschingetal.2020}). Our model also suggests a higher benefit of the control treatment in patients with progesterone receptor–positive (PR+) tumors compared to PR$-$ tumors (point estimate $-$0.0542 [$-$0.0787; $-$0.0298]). Across tumor stage categories (pT1–pT4), the estimated coefficients indicate an increasing trend, with the highest estimated benefit for pT4 tumors compared to pT1 tumors (0.1324 [0.0336; 0.2312]). In patients with invasive lobular tumors or other invasive epithelial breast cancers, our model suggests a higher DFS advantage for the control group compared to invasive ductal breast cancers; note, however, that the 95\% confidence interval for invasive lobular tumors includes the value zero. This observation aligns with the trend reported in the original publication \citep{GregorioHaeberleFaschingetal.2020}.

Our analysis also highlights some of the conceptual differences between the proposed model (Figure \ref{fig:dataexample}A) and the Cox proportional hazards model (Figure \ref{fig:dataexample}B): First, as shown in Appendix~\ref{app:Cox}, the group-by-covariate interaction effects in Cox regression (presented in Figure \ref{fig:dataexample}B) have a similar interpretation as the effects ${\bs\beta}_1 + {\bs\beta}_2$ in the proposed model with logistic link function, provided that the Cox model is correctly specified and $\tau = \infty$. It follows that the estimates shown in Figure~\ref{fig:dataexample}A refer to covariate effects on the original (probability) scale, whereas the Cox-based estimates in Figure \ref{fig:dataexample}B have to be interpreted on the logit scale. We consider this improved interpretability (combined with the general flexibility in the choice of the link function) as a major strength of our model, in particular since all estimated conditional Mann-Whitney effects keep the probability range~$[0,1]$ (cf.\@ Figure \ref{fig:dataexample}C). Second, there is a conceptual difference between the proposed model and the Cox model in the definition of the underlying estimands: As stated above, the  coefficients of the proposed model refer to the ``truncated'' conditional Mann-Whitney effect $P(\min(T_1, \tau) > \min(T_2, \tau) \mid \bs Z)$ with finite time horizon $\tau = 5.5$ years. In contrast, as shown in Appendix \ref{app:Cox}, the coefficients of the Cox model have a slightly different interpretation, as they translate to the {\em untruncated} conditional Mann-Whitney effect $P(T_1 > T_2 \mid \bs Z)$ with $\tau = \infty$. (Note that admininstrative censoring at~5.5~years does not violate the independent censoring assumption; consequently, the partial likelihood estimator remains consistent for the main and interaction effects of the Cox model.) This difference in estimands is most likely the reason for some of the changes in the effect direction when comparing Figures~\ref{fig:dataexample}A and~B (cf., in particular, the positive effects of tumor stages in Figure \ref{fig:dataexample}A and the negative effects of tumor stages in Figure~\ref{fig:dataexample}B). It is, in fact, easy to reproduce such sign flips in simulated data, see {\color{black}the file \emph{Illustrative\_example\_Cox\_model.R} at \emph{https://github.com/dennis-dobler/Mann-Whitney\_Regression}.} Alternatively, the observed differences between the effect directions in Figures~\ref{fig:dataexample}A and~\ref{fig:dataexample}B could have been caused by violations of the proportional hazards assumption.

Figure \ref{fig:dataexample}C presents the estimated tie-corrected truncated Mann-Whitney effects, as obtained from Eq.\@ \eqref{eq:predictions} for all patients in the data set ($n =$ 3,652). Based on these estimated probabilities and their 95\% confidence intervals, patients were classified as follows: Patients with a lower bound of the 95\% confidence interval larger than 0.5 were classified as benefiting from the intervention, whereas patients with an upper bound of the 95\% confidence interval  
smaller than 0.5 were classified as benefiting from the control treatment. Overall, 1,112 (30.4\%) patients were estimated to benefit from being in the control group (orange lines in Figure \ref{fig:dataexample}C), whereas 1,311 (35.9\%) were estimated to benefit from the intervention (turquoise lines in Figure~\ref{fig:dataexample}C), illustrating the heterogeneity of treatment effects across the patients. Notably, patients estimated to benefit from the intervention showed a markedly higher prevalence of advanced disease characteristics, including higher tumor stage (pT2–pT4, 76.1\%) and poor differentiation (G3, 73.7\%, see Table \ref{tab:supplement_SUCCESS_A_benefit_harm}). Hormone receptor statuses differed substantially between groups: Estrogen and progesterone receptor positivity were common among patients estimated to benefit from the control treatment (ER+: 87.7\%, PR+: 82.6\%) but less frequent among those estimated to benefit from the intervention (ER+: 44.9\%, PR+: 33.5\%, see Table~\ref{tab:supplement_SUCCESS_A_benefit_harm}). 

Figure \ref{fig:dataexample}D illustrates the estimated tie-corrected truncated Mann-Whitney effects, as obtained from Eq.\@ \eqref{eq:predictions}, across patient subgroups defined by molecular tumor subtypes \citep{Perou2000}. Here, {\em HER2-positive} patients are characterized by HER2-positive tumors, irrespective of estrogen receptor (ER) status, progesterone receptor (PR) status, or tumor grade. HER2-negative tumors are further subdivided into {\em luminal A–like} tumors (HER2-negative, ER and/or PR positive, grade G1 or G2), {\em luminal B–like} tumors (HER2-negative, ER and/or PR positive, grade G3), and {\em triple-negative} tumors (HER2-, ER-, and PR-negative, any grade). As shown in Figure \ref{fig:dataexample}D, more than 75\% of patients with triple-negative tumors exhibit an estimated Mann-Whitney effect exceeding 0.5. Accordingly, our model suggests that these patients tend to have a higher benefit (in terms of DFS) from the intervention than from the control therapy. Further subgroup-specific estimates are presented in Figure \ref{fig:supp_predictions}.

\section{Discussion}
\label{sec:disc}

We presented a statistical model for the two-sample problem that relates the {\color{black}conditional} Mann-Whitney effect $\theta ({\bs z_1}, {\bs z_2}) = P(T_1 > T_2 | {\bs Z_1} = {\bs z_1}, {\bs Z_2} = {\bs z_2})$ to group-specific covariate values ${\bs z_1}, {\bs z_2}$. The proposed model is distribution-free, thereby generalizing nonparametric estimators of the unconditional Mann-Whitney effect (like the Wilcoxon-Mann-Whitney statistic) to a regression setting. {\color{black}If the logistic link function is used, our model has a similar interpretation as the conditional Mann-Whitney effect in a Cox regression model with group-covariate interactions.} As a main result of this paper, we developed a model-fitting procedure based on pseudo-observations. We showed that this procedure yields a consistent and asymptotically normal estimator of the covariate effects under rather mild conditions. In the special case of uncensored observations, we explicitly verified these conditions. In future work, we will investigate more detailed sufficient conditions for the validity of Assumption~\ref{ass:general} in the randomly right-censored case.

In addition to the aforementioned model-fitting procedure, we proposed several bootstrap-based statistical hypothesis tests to infer possible covariate effects on $\theta$. Compared to other non-parametric inference procedures for the two-sample problem (like the log-rank test), our model has the advantage that it additionally yields {\em effect estimates} in terms of unconditional and conditional probabilities $P(T_1 > T_2)$ and $P(T_1 > T_2 | {\bs Z_1} = {\bs z_1}, {\bs Z_2} = {\bs z_2})$, respectively. As demonstrated by our analysis of the SUCCESS-A study data, this estimation and testing approach could be used to identify covariates that are predictive for the treatment effect, corresponding to a simultaneous analysis of subgroups in a clinical trial. Our approach further yields patient-specific probability estimates with confidence intervals, allowing for a classification scheme to identify those patients having the highest expected benefit from a specific treatment (cf.\@ Section \ref{sec:data}).

An attractive feature of the proposed model is its ability to deal with different sets of covariate values, and even different sets of covariates, across the two sample groups. As a result, our model allows for very flexible two-sample comparisons, e.g., when intervention and control data have been obtained from different sources (as in the case of synthetic and/or historical control arms, possibly involving different sets of covariates).
It also has a direct connection to causal inference: For the following discussion, let us for simplicity consider the case  of continuously distributed outcome variables.
As \cite{fay} argue, it is possible to write $P(T_{1i_1} > T_{2i_2})$ as a ``quantile difference causal effect''. To understand this quantity a bit better and using the same notation ($\Phi, \Psi$) as \cite{fay}, we rewrite the Mann-Whitney effect as
\begin{align*}
    \Phi & = P(T_{1i_1} > T_{2i_2}) \\
    & = \int \int P(T_{1i_1} > T_{2i_2} \  | \  {\bs Z_{1i_1}} = {\bs z_1}, {\bs Z_{2i_2}} = {\bs z_2}) d P^{\bs Z}(\bs z_1) \, dP^{\bs Z}(\bs z_2).
\end{align*}
Here, $P^{\bs Z}$ stands for a target population's covariate distribution for which causal inference shall be drawn.
However, this quantity might suffer from the so-called \textit{Hand's paradox}: In the case of matched pairs in both treatment groups, e.g., even in identical twin studies with i.i.d.\ pairs of event times and \textcolor{black}{covariable vectors} $(\tilde T_{1i}, \tilde T_{2i}, \textcolor{black}{\bs Z_i})$, $i=1,\dots, n$, \textcolor{black}{where the subscripts 1 and 2 indicate the administration of the intervention and control treatments, respectively. Here}, it is possible that for a majority of pairs $\tilde T_{1i} > \tilde T_{2i}$ holds true although $P(\tilde T_{1i_1} > \tilde T_{2i_2}) < 0.5$ for $i_1 \neq i_2$. This phenomenon is alleviated if we instead consider
\begin{align*}
    \tilde \theta & = P(T_{1i_1} > T_{2i_2} \ | \ {\bs Z_{1i_1}} = {\bs Z_{2i_2}} ) \\
    & = \int P(T_{1i_1} > T_{2i_2} \ | \ {\bs Z_{1i_1}} = {\bs z}, {\bs Z_{2i_2}} = {\bs z}) \, d P^{\bs Z}(\bs z) \\
    & = - \int \int S_1(t \ | \  {\bs z})  S_2(d t \ | \  {\bs z}) \, d P^{\bs Z}(\bs z).
\end{align*}
In identical twin studies, this quantity is as close as one can get to the actual counterfactual framework where one would instead want to estimate
$$ \Psi = P(\tilde T_{1i} > \tilde T_{2i}) = \int P(\tilde T_{1i} > \tilde T_{2i} \ | \ {\bs Z_{i}} = {\bs z}) \, d P^{\bs Z}(\bs z), $$
In most cases of two-sample problems, $\Psi$ stands for an ideal one cannot hope to estimate, \textcolor{black}{e.g., in counterfactual settings for causal analyses.}
However, in perfectly matched pairs studies, if sufficient covariate information is available, one would expect that $\tilde \theta$ is closer to the causal estimand $\Psi$ than to $\Phi$.
Indeed, if one has identified and measured \textit{all} covariates associated with the outcome---the assumption of no unmeasured confounding is of course often unrealistic---then one may argue that $\tilde \theta \approx \Psi$.
Specifically, based on the present regression approach, estimation of the unconditional estimand $\tilde \theta$ should be possible under the usual consistency, positivity, and exchangeability conditions. For example, in the SUCCESS-A study, using a g-computation approach, an unconditional contrast could be obtained a weighted average of the patient-specific probabilities in Figure \ref{fig:dataexample}C.

From a methodological perspective, it seems possible to extend the  two-sample pseudo-observation-based regression approach to other functionals of the marginal distributions, $\theta= \phi(S_1, S_2)$, as long as there exists a sensible estimator, $\hat \theta= \phi(\hat S_1, \hat S_2)$.
To verify the statistical properties of the resulting estimators in such a general approach, it would be necessary to impose additional assumptions, e.g., a certain type of differentiability and that the estimand can be expressed as a conditional expectation; 
see \cite{ovpape17} for the large-sample behavior in the one-sample case. 

Extensions of the proposed jackknife-based pseudo-observation regression approach to $J$-sample problems, $J>2$ are more or less straightforward: This would involve a $J$-way array of pseudo-observations where each possible combination of $J$-tuples are left-out exactly once, resulting in $N := \prod_{j=1}^J n_j$ pseudo-observations in total.
The pseudo-observations related to a $J$-sample estimand $\theta$ would then take the form
\begin{align*}
 \tilde \theta_{i_1 \dots i_J}
 = \ &  N \cdot \Big[ \hat \theta - \frac{n_1-1}{n_1} \,\hat \theta_1^{(i_1)} - \frac{n_2-1}{n_2} \,\hat \theta_2^{(i_2)} - \dots - \frac{n_J-1}{n_J} \,\hat \theta_J^{(i_J)} \\
 & + \frac{(n_1-1)(n_2-1)}{n_1 n_2} \,\hat \theta_{12}^{(i_1 i_2)} + \frac{(n_1-1)(n_3-1)}{n_1 n_3} \,\hat \theta_{13}^{(i_1 i_3)} + \dots  \\
 & - \frac{(n_1-1)(n_2-1)(n_3-1)}{n_1 n_2 n_3} \,\hat \theta_{123}^{(i_1 i_2 i_3)} - \dots \\
 & +  \dots  + (-1)^J \, \Big(\prod_{j=1}^J \frac{n_j-1}{n_j}\Big) \,\hat \theta^{(i_1 \dots i_J)}_{1\dots J}\Big] \\
 =: \ & N  \cdot \sum_{I \subset \{ 1, \dots, J\}} (-1)^{|I|} \,\Big(\prod_{k \in I} \frac{n_k-1}{n_k}\Big)\, \hat \theta_{I}^{(i_I)} ,
\end{align*}
where, in the last step, we used the obvious multi-index notation with subscripts indicating omission of the individual observations  $i_I$
from the set of sample indices $I$.
Here, $\hat \theta = \hat \theta_\emptyset^{(i_\emptyset)}$ denotes the full-sample(s)-based estimator.

\section*{Conflicts of interest}
The authors declare that they have no competing interests.


\newpage
\section*{Data availability}
{\color{black}The code used for the simulation study in Section~\ref{sec:simus} is accessible at \emph{https://github.com/dennis-dobler/Mann-Whitney\_Regression}. This repository also contains an example on how to compare the proposed model to a Cox model with group-covariate interactions, and on how to fit the proposed model using other link functions than the identity function. The SUCCESS-A study data are not publicly available due to
privacy and ethical restrictions.}

\section*{Author contributions statement}

D.D.\ suggested a first draft of the methodological parts, derived the mathematical-statistical results and proofs, wrote the R code {\color{black}for the} implementation of the method, and ran the simulation study.
A.S.\ prepared all figures and reviewed and adjusted the R code as needed for the analysis of the SUCCESS-A data.
The data analysis was conducted by A.S.\ and M.S.
M.S.\ prepared the R code for the geese-based implementation of the regression for the Mann-Whitney effect and the comparison with the Cox model, and made conceptual and editorial changes and additions to all parts of the paper.
D.D., A.S., and M.S.\ jointly designed the paper structure, interpreted all empirical results, and wrote and reviewed the manuscript.

\section*{Acknowledgments}
The authors thank Lukas Nagy (TU Dortmund University, Germany) for his support in implementing and exploring part of the method in R, and Simon Mack (RWTH Aachen University, Germany), Erik T.\ Parner and Morten Overgaard (both Aarhus University, Denmark), Edgar Brunner (University Medical Center Göttingen, Germany), and the late Marc Ditzhaus (Otto von Guericke University Magdeburg, Germany) for helpful discussions. The authors also thank Lothar Häberle (Department of Gynecology and Obstetrics, Erlangen University Hospital, Germany) for his support with the analysis of the SUCCESS-A study data.


\newpage

\appendix
\section*{Appendix}

\section{Mathematical proofs}
\label{app:proofs}

\textbf{Proof of the central limit theorem in Example~\ref{ex:identity}.}
We first note that the expectation of $\hat {\bs\Sigma}$ equals
$$ \bs \Sigma = \text{E} \Big(\Big(\begin{smallmatrix} 1 \\ \bs Z_{11} \\ \bs Z_{21} \end{smallmatrix}\Big)\Big(\begin{smallmatrix} 1 \\ \bs Z_{11} \\ \bs Z_{21} \end{smallmatrix} \Big)^\top \Big). $$
By the law of large numbers applied to each of the two samples, it is easy to see that $\bs \Sigma$ is the almost sure limit of $\hat {\bs \Sigma}$ (since both $\bs Z_{1i}$ and $\bs Z_{2j}$ have finite second moments by assumption).

Similarly, the expectation of $\hat{\bs \Psi}$ equals
$$ \bs \Psi = \text{E} \Big( \Big(\begin{smallmatrix} 1 \\ \bs Z_{11} \\ \bs Z_{21} \end{smallmatrix}\Big) 1\{T_{11} > T_{21} \} \Big) , $$
which could be slightly simplified using the independence of the two samples.
Also note that the first entry equals the unconditional Mann-Whitney effect $\theta$. By definition, $\hat{\bs \Psi}$ has the structure of a $U$-statistic which is similar to the Mann-Whitney $U$-statistic.
As a consequence, classical theory for $U$-statistics (e.g., Theorem~4.5.1 of \citealp{kobo94}) in combination with Slutsky's theorem 
implies that
$$\bs  \Omega_{n_1n_2}^{-1/2}(\hat{\bs \Psi} - \bs \Psi) \stackrel d \to \mathcal N_p(\bs 0, \bs I_p) , $$
provided that the matrix
\begin{align}
\begin{split}
\label{eq:omega}
 \bs \Omega_{n_1n_2} \, & = \, a_{n_1n_2} \bs \Omega^{(0)} + b_{n_1n_2} \bs \Omega^{(1)} + c_{n_1n_2} \bs \Omega^{(2)} \\
  & := \frac1{n_1n_2} 
 \text{var}\Big(\Big(\begin{smallmatrix} 1 \\ \bs Z_{11} \\ \bs Z_{21} \end{smallmatrix}\Big) 1\{T_{11} > T_{21} \} \Big) \\
 & \phantom{:=} + \frac{n_2-1}{n_1n_2} \, \text{cov}\Big(\Big(\begin{smallmatrix} 1 \\ \bs Z_{11} \\ \bs Z_{21} \end{smallmatrix}\Big) 1\{T_{11} > T_{21} \}, \Big(\begin{smallmatrix} 1 \\ \bs Z_{11} \\ \bs Z_{22} \end{smallmatrix}\Big) 1\{T_{11} > T_{22} \} \Big) \\
 & \phantom{:=} + \frac{n_1-1}{n_1n_2} \, \text{cov}\Big(\Big(\begin{smallmatrix} 1 \\ \bs Z_{11} \\ \bs Z_{21} \end{smallmatrix}\Big) 1\{T_{11} > T_{21} \}, \Big(\begin{smallmatrix} 1 \\ \bs Z_{12} \\ \bs Z_{21} \end{smallmatrix}\Big) 1\{T_{12} > T_{21} \} \Big) 
\end{split}
\end{align}
has full rank. Here, $\bs \Omega_{n_1n_2}^{-1/2} := (\bs \Omega_{n_1n_2}^{1/2})^{-1}$, where the square root is defined by the positive definite matrix satisfying $\bs \Omega_{n_1n_2}^{1/2} \bs \Omega_{n_1n_2}^{1/2} = \bs \Omega_{n_1n_2}$. Finally, using $n_1/(n_1+n_2) \to \lambda \in (0,1)$
as well as 
$$ \sqrt{\frac{n_1 n_2}{n_1 + n_2}} \, \hat {\bs \Sigma}^{-1}\hat {\bs \Psi} = \hat {\bs \Sigma}^{-1} \Big( \frac{n_1 n_2}{n_1 + n_2} \, \bs\Omega_{n_1n_2}\Big)^{1/2} \Big(\bs\Omega_{n_1n_2}^{-1/2} \hat {\bs \Psi}  \Big), $$
the previous results combined with Slutsky's theorem yield
$$ \sqrt{\frac{n_1 n_2}{n_1 + n_2}} \, (\hat {\bs \beta} - \bs \beta_0) = \sqrt{\frac{n_1 n_2}{n_1 + n_2}} \, (\hat {\bs \Sigma}^{-1}\hat {\bs \Psi} - \bs \Sigma^{-1}\bs \Psi) \, \stackrel d \to \, \mathcal N_p(\bs 0, \bs \Sigma^{-1} \bs \Omega (\bs \Sigma^{-1})^\top) $$
as $\min(n_1,n_2) \to \infty$, where $\bs \Omega = (1-\lambda) \bs \Omega^{(1)} + \lambda \bs \Omega^{(2)}$. \hfill \qed \\

Estimates of the matrices $\hat{\bs \Omega}^{(0)}$, $\hat{\bs \Omega}^{(1)}$ and $\hat{\bs \Omega}^{(2)}$  can be obtained by using the empirical estimators
\begin{align*}
     \hat{\bs \Omega}^{(0)} & = 
     \frac{1}{n_1n_2} 
     \sum_{i_1=1}^{n_1}
     \sum_{i_2=1}^{n_2}
     1\{T_{1i_1} > T_{2i_2} \}
     \Big(\begin{smallmatrix} 1 \\ \bs Z_{1i_1} \\ \bs Z_{2i_2} \end{smallmatrix}\Big)  \Big(\begin{smallmatrix} 1 \\ \bs Z_{1\tilde i_1} \\ \bs Z_{2 \tilde i_2} \end{smallmatrix}\Big)^\top  
      \\
      & \quad - 
     \Big( \frac{1}{n_1n_2} 
     \sum_{i_1=1}^{n_1}
     \sum_{i_2=1}^{n_2}
     1\{T_{1i_1} > T_{2i_2} \}
     \Big(\begin{smallmatrix} 1 \\ \bs Z_{1i_1} \\ \bs Z_{2i_2} \end{smallmatrix}\Big) \Big) 
     \Big( \frac{1}{n_1n_2} 
     \sum_{i_1=1}^{n_1}
     \sum_{i_2=1}^{n_2}
     1\{T_{1i_1} > T_{2i_2} \}
     \Big(\begin{smallmatrix} 1 \\ \bs Z_{1i_1} \\ \bs Z_{2i_2} \end{smallmatrix}\Big)  \Big)^\top , \\
     \hat{\bs \Omega}^{(1)} & = \frac{1}{n_1n_2^2} 
     \sum_{i_1=1}^{n_1}
     \sum_{i_2=1}^{n_2}\sum_{\tilde i_2=1}^{n_2}
1\{T_{1  i_1} > T_{2 \tilde i_2} \}
     1\{T_{1i_1} > T_{2i_2} \}
     \Big(\begin{smallmatrix} 1 \\ \bs Z_{1i_1} \\ \bs Z_{2i_2} \end{smallmatrix}\Big)  \Big(\begin{smallmatrix} 1 \\ \bs Z_{1 i_1} \\ \bs Z_{2 \tilde i_2} \end{smallmatrix}\Big)^\top   \\
      & \quad - 
     \Big( \frac{1}{n_1n_2} 
     \sum_{i_1=1}^{n_1}
     \sum_{i_2=1}^{n_2}
     1\{T_{1i_1} > T_{2i_2} \}
     \Big(\begin{smallmatrix} 1 \\ \bs Z_{1i_1} \\ \bs Z_{2i_2} \end{smallmatrix}\Big) \Big) 
     \Big( \frac{1}{n_1n_2} 
     \sum_{i_1=1}^{n_1}
     \sum_{i_2=1}^{n_2}
     1\{T_{1i_1} > T_{2i_2} \}
     \Big(\begin{smallmatrix} 1 \\ \bs Z_{1i_1} \\ \bs Z_{2i_2} \end{smallmatrix}\Big)  \Big)^\top , \ \ {\text{and}}\\
     \hat{\bs \Omega}^{(2)} & = \frac{1}{n_1^2n_2} 
     \sum_{i_1=1}^{n_1}\sum_{\tilde i_1=1}^{n_1} 
     \sum_{i_2=1}^{n_2}
     1\{T_{1i_1} > T_{2i_2} \}  1\{T_{1 \tilde i_1} > T_{2  i_2} \}
     \Big(\begin{smallmatrix} 1 \\ \bs Z_{1i_1} \\ \bs Z_{2i_2} \end{smallmatrix}\Big) \Big(\begin{smallmatrix} 1 \\ \bs Z_{1\tilde i_1} \\ \bs Z_{2  i_2} \end{smallmatrix}\Big)^\top   \\
      & \quad - 
     \Big( \frac{1}{n_1n_2} 
     \sum_{i_1=1}^{n_1}
     \sum_{i_2=1}^{n_2}
     1\{T_{1i_1} > T_{2i_2} \}
     \Big(\begin{smallmatrix} 1 \\ \bs Z_{1i_1} \\ \bs Z_{2i_2} \end{smallmatrix}\Big) \Big) 
     \Big( \frac{1}{n_1n_2} 
     \sum_{i_1=1}^{n_1}
     \sum_{i_2=1}^{n_2}
     1\{T_{1i_1} > T_{2i_2} \}
     \Big(\begin{smallmatrix} 1 \\ \bs Z_{1i_1} \\ \bs Z_{2i_2} \end{smallmatrix}\Big)  \Big)^\top ,
\end{align*}
respectively.

\bigskip
\noindent
\textbf{Proof of Theorem~\ref{thm:censored}:}
Since $- \check{\bs \Sigma}(\bs \beta)$ is negative definite and continuous in $\bs \beta_0$, it is also negative definite on the open ball $B = B_\varepsilon(\bs \beta_0)$ of some radius $\varepsilon > 0$ around $\bs \beta_0$. This is because the largest eigenvalue of $- \check{\bs \Sigma}(\bs \beta)$ changes continuously with $\bs \beta$. Let $\varepsilon>0$ be small enough so that the largest eigenvalue of all $- \check{\bs \Sigma}(\bs \beta)$, $\bs \beta \in B$, is bounded away from~0, and such that the assumed uniform convergence of $D\bs U$ to $-\check{\bs \Sigma}$ in probability holds on $B$.
As a consequence, the event that the biggest eigenvalue of $D\bs U$ is negative has a probability converging to 1 as $\min(n_1,n_2)\to \infty$. Hence it is no restriction to consider all arguments in the remainder of this proof conditionally on the event
$$D_{n_1n_2}(\varepsilon) := \{D\bs U (\bs \beta) \text{ is negative definite for all } \bs \beta \in B_{\varepsilon}(\bs \beta_0)\}.$$

Using the arguments from \cite{angi82}, Appendix~II, the maximum of $\bs \beta \mapsto C(\bs \beta)$ (over the compact subset $\overline{B_{\varepsilon/2}(\bs \beta_0)}$, i.e., the closed ball) converges to the maximizing value of $C$ in probability, i.e., to $\bs \beta_0$.
 Now Corollary~II.2 of \cite{angi82} implies that the maximizer $\hat {\bs \beta}$ of $C$ eventually lies in the open set $B_{\varepsilon/2}(\bs \beta_0)$ in probability, i.e., it is a stationary point of $\bs U$. Consequently, the probability that the root of $\bs U$ exists tends to 1.
For these arguments we also used that $\bs \beta_0$ is the only root of $\nabla f$ and that $C$ converges pointwise in probability to $f$. In combination with Corollary~II.2 of \cite{angi82}, it follows from $\bs \beta_0$ being the only stationary point of $f$ that $\hat{\bs \beta}$ exists with a probability tending to~1.

Now, applying the mean value theorem individually to each component $U_k(\bs \beta)$, $k=1, \dots, p$, of $\bs U(\bs \beta)$, there exist points $\bs \beta_k$ on the line segment between $\bs \beta_0$ and $\hat{\bs \beta}$ such that 
$$U_k(\hat{\bs \beta}) - U_k(\bs \beta_0) = \nabla U_k(\bs \beta_k) (\hat{\bs \beta} - \bs \beta_0). $$
Denote by $\widetilde {D \bs U}(\hat{\bs \beta}, \bs \beta_0) $ the matrix consisting of the rows $\nabla U_k(\bs \beta_k), k=1, \dots, p$.
Then, by the existence of the root $\hat{\bs \beta}$ of $\bs U$, it follows that 
$$ - \bs U(\bs \beta_0) = \widetilde{D \bs U} (\hat{\bs \beta}, \bs \beta_0) \, (\hat{\bs \beta} - \bs \beta_0). $$
The matrix $\widetilde{D \bs U}(\hat{\bs \beta}, \bs \beta_0)$ is invertible due to conditioning on the set $D_{n_1n_2}(\varepsilon)$.
The uniform convergence of $D \bs U$ to $- \check{ \bs \Sigma} $ in probability over $B_{\varepsilon/2}(\bs \beta_0)$ and the continuity of the limit in $\bs \beta_0$ yield 
$\widetilde{D \bs U}(\hat{\bs \beta}, \bs \beta_0) \stackrel p \to - \check{ \bs \Sigma}(\bs \beta_0)$.
Combining these arguments, using the assumed central limit theorem for $\bs U(\bs \beta_0)$ postulated in Assumption~\ref{ass:general}.5, and using Slutsky's theorem, we obtain
\begin{align*}
    \sqrt{\frac{n_1n_2}{n_1+n_2}}\, (\hat{\bs \beta} - \bs \beta_0) \, = \, - \sqrt{\frac{n_1n_2}{n_1+n_2}} \, (\widetilde {D \, \bs U}(\hat{\bs \beta}, \bs \beta_0) )^{-1} \bs U(\bs \beta_0) \, \stackrel d \to \, \check{\bs \Sigma}(\bs \beta_0) \bs W
\end{align*}
as $\min(n_1,n_2) \to \infty$, with the claimed multivariate normal distribution.
\hfill \qed


\section{Additional information about the hazard rates in the simulation study}
\label{app:simu}

In case of Weibull-distributed event times, the Mann-Whitney effect simplifies considerably if both shape parameters coincide, i.e., $k_1=k_2=k$. It is, in fact, easy to see that, for the Mann-Whitney effect truncated at $\tau$,
\begin{align*}
\theta &  = P(\min(T_1,\tau) > \min(T_2,\tau)) = (1 - S_1(\tau)S_2(\tau))\, \frac{\lambda_1^k}{\lambda_1^k + \lambda_2^k} \\
& =  \frac{1 - \exp(-\tau^k (\exp(-k\gamma_{10}-k \bs \gamma_1^\top \bs Z_1) + \exp(-k\gamma_{20}-k \bs \gamma_2^\top\bs  Z_2) ))}{1+ \exp(-k(\gamma_{10} + \bs \gamma_1^\top\bs  Z_1 - \gamma_{20} - \bs \gamma_2^\top\bs  Z_2))} . 
\end{align*}
Furthermore, the parameter vectors $\bs \gamma_j$ have an interpretation similar to $\bs \beta_j$, $j=1,2$.
This is best seen for large $\tau$ for which
\begin{equation}
\label{weibHazard}
\theta \,\approx \, \frac1{1+ \exp(-k(\gamma_{10} + \bs \gamma_1^\top \bs Z_1 - \gamma_{20} - \bs \gamma_2^\top \bs Z_2))} .
\end{equation}
Note, however, that the signs of the groups' parameter vectors in \eqref{weibHazard} are different. Eq.\@ (\ref{weibHazard}) also reveals that the proposed model approaches a logistic regression model with $\beta_0 = k \cdot (\gamma_{10} - \gamma_{20})$, $\bs \beta_1 = k \cdot \bs \gamma_1 $, and $\bs \beta_2 = -k \cdot \bs \gamma_2$ if $k_1=k_2=k$ and $\tau$ is large.
We also refer to Appendix~\ref{app:Cox} below for the connection between the Mann-Whitney effect and the Cox model in the case of the logistic link function.

In the special case of no covariate effects, $\bs \gamma_j\equiv \bs 0$, $j=1,2$, which could be of interest for hypothesis testing,
all regression models are correctly specified, and it holds that $\bs \beta_1 = \bs \beta_2= \bs 0$.
In all other cases and for other common choices of the link function, 
the model is misspecified. Nevertheless, testing hypotheses about zero covariate effects remains meaningful, cf.\ Table~\ref{tab:gamma_parms}. If the shape parameters are unequal, i.e.\@ $k_1\neq k_2$, there is no such simple representation of $\theta$.

\newpage

\section{Derivation of the Mann-Whitney effect in  Cox regression}
\label{app:Cox}

Here, we derive an explicit formula for the Mann-Whitney effect ($\tau = \infty$) under the assumptions of a Cox regression model. In addition, we demonstrate how changes in a covariate translate to changes in the Mann-Whitney effect in the absence of competing events. To this end, we consider a Cox regression model of the form
\begin{equation}
\label{Cox1}
\lambda (t|g_{s}, \bs z_{s}) = \lambda_0 (t) \cdot \exp (\eta_0 \cdot g_{s} + \eta_1 \cdot z_{1s} + \ldots + \eta_{p} \cdot z_{p s} + \delta_1 \cdot g_{s} \cdot z_{1s} + \ldots + \delta_{p} \cdot g_{s} \cdot z_{ps}) =:  \lambda_0 (t) \cdot \exp (\xi_s),
\end{equation}
where $\lambda (\cdot)$ and $\lambda_0 (\cdot)$ denote the hazard and baseline hazard functions, respectively, and $g_{s} \in \{0,1\}$ and $\bs z_s = (z_{1s},\ldots , z_{ps})^\top$, $s = 1, \ldots , n_1+n_2$, are the sample values of the binary group indicator  and the $p = p_1 = p_2$ covariates, respectively. Correspondingly, $\eta_0$ and $\eta_1, \ldots , \eta_{p}$ are the main effects of the group indicator and the covariates, respectively. The group-covariate interaction effects are denoted by $\delta_1, \ldots , \delta_{p}$. The group indicator takes the value $g_s=1$ if individual $s$ belongs to sample group 1 and $g_s=0$ if $s$ belongs to sample group 2. We fitted model \eqref{Cox1} to all simulated data sets in Section \ref{sec:simus}, and also to the SUCCESS-A study data in Section \ref{sec:data}.

Let $s_1$ and $s_2$ be two randomly chosen individuals from sample groups 1 and 2, respectively. Under the assumptions of model \eqref{Cox1}, the Mann-Whitney effect is derived as
\begin{eqnarray}
\theta (\bs z_{s_1} , \bs z_{s_2}) \, = \, P(T_1 > T_2 | \bs z_{s_1} , \bs z_{s_2}) &=& - \int_0^\infty S_1(t| \bs z_{s_1}) \, d S_2(t| \bs z_{s_2}) \nonumber\\
&=& \int_0^\infty S_1(t| \bs z_{s_1}) \cdot S_2(t| \bs z_{s_2}) \cdot \lambda (t| 0, \bs z_{s_2}) \, dt \nonumber\\
&=& \int_0^\infty S_0 (t)^{\exp (\xi_{s_1})} \cdot S_0 (t)^{\exp (\xi_{s_2})} \cdot \lambda_0 (t) \cdot \exp (\xi_{s_2}) \, dt ,
\end{eqnarray}
where $S_0 (t) := \exp (- \int_0^t \lambda_0 (u)du)$ denotes the baseline survival function of the Cox regression model. It follows that
\begin{eqnarray}
\label{relEffectCox}
\theta (\bs z_{s_1} , \bs z_{s_2}) 
&=& \int_0^\infty S_0 (t)^{\exp (\xi_{s_1}) + \exp(\xi_{s_2})} \cdot \lambda_0 (t) \cdot \exp (\xi_{s_2}) \, dt \nonumber\\
&=& \left[ - \frac{\exp (\xi_{s_2})}{\exp (\xi_{s_1}) + \exp(\xi_{s_2})} \cdot S_0 (t)^{\exp (\xi_{s_1}) + \exp(\xi_{s_2})}  \right]_0^\infty \nonumber\\
&=& \frac{1}{1 + \exp (\xi_{s_1}-\xi_{s_2})} \ = \ \frac{1}{1 + \exp (- (\xi_{s_2}-\xi_{s_1}))} ,
\end{eqnarray}
see also \cite{nege20} for a similar relationship in the probabilistic index model under proportional hazards.
Additionally, \cite{oakes16} pointed out that the hazard ratio equals the so-called loss ratio in the Cox model; 
the loss ratio is defined as the Mann-Whitney effect divided by one minus the Mann-Whitney effect. In particular, the relationship in (\ref{relEffectCox}) is in line with Eq.~(\ref{weibHazard}) above, since the pairs of Weibull models considered in Appendix \ref{app:simu} are equivalent to a Cox regression model if $k=k_1=k_2$.

We are interested in how changes in a specific covariate $Z^*$ affect the Mann-Whitney effect in \eqref{relEffectCox}. Without loss of generality, we assume that $Z^*$ is the first of the $p$ covariates. In this case, one can rewrite the linear combination $\xi_{s_2}-\xi_{s_1}$ as
\begin{eqnarray}\label{xi1xi2}
\xi_{s_2}-\xi_{s_1} &=& -\eta_0 + \eta_1 \cdot (z^*_{s_2}- z^*_{s_1}) - \delta_1 \cdot z^*_{s_1} + \eta_2 \cdot (z_{2s_2}-z_{2s_1}) + \ldots \nonumber \\
&&+ \eta_{p} \cdot (z_{ps_2}-z_{ps_1}) - \delta_2 \cdot z_{2s_1} - \ldots - \delta_{p} \cdot z_{ps_1} \nonumber\\
&=& -\eta_0 - (\eta_1+\delta_1) \cdot z^*_{s_1} + \eta_1 \cdot z^*_{s_2} + \mathcal{R}(z_{2s_1}, \ldots , z_{ps_1}, z_{2s_2}, \ldots , z_{ps_2} ) ,
\end{eqnarray}
where $z^*_{s_1},z^*_{s_2}$ denote the sample values of $Z^*$ and $\mathcal{R}$ is a remainder term defined by $$\mathcal{R} \equiv \mathcal{R}(z_{2s_1}, \ldots , z_{ps_1}, z_{2s_2}, \ldots , z_{ps_2} ) := \xi_{s_2}-\xi_{s_1} + \eta_0 - \eta_1 \cdot (z^*_{s_2}-z^*_{s_1}) + \delta_1 \cdot z^*_{s_1}.$$

From \eqref{relEffectCox} and \eqref{xi1xi2} it is seen that the Mann-Whitney effect in a Cox regression model corresponds to the definition of the Mann-Whitney effect in the proposed model (\ref{eq:model2}) with logistic link function, i.e.,  $\mu (x) = 1/(1+\exp(-x))$. In this case, the interpretation of $ \xi_{s_2}-\xi_{s_1}$ is analogous to the interpretation of the term $
\beta_0 + \bs \beta_1^\top \bs z_1 + \bs \beta_2^\top\bs  z_2$ in Eq.~(\ref{eq:model2}). 
In particular, testing for ``$H_0^{(1)}: \beta_{11}=0$ vs.\@ $H_1^{(1)}: \beta_{11}\neq0$'' and ``$H_0^{(2)}: \beta_{21}=0$ vs.\@ $H_1^{(2)}: \beta_{21}\neq0$'' corresponds to testing for ``$H_{0,\text{Cox}}^{(1)}: -(\eta_{1}+\delta_1)=0$ vs.\@ $H_{1,\text{Cox}}^{(1)}: -(\eta_{1}+\delta_1)\neq0$'' and ``$H_{0,\text{Cox}}^{(2)}: \eta_1=0$ vs.\@ $H_{1,\text{Cox}}^{(2)}: \eta_1\neq0$'', respectively, as done in our simulation study in~Section~\ref{sec:simus}.

In our analysis of the SUCCESS-A study, we only consider the special case $\bs Z_1 = \bs Z_2 = \bs Z$, implying $z^*_{s_1}=z^*_{s_2} = z^*$ in Eq.~\eqref{xi1xi2}. In this case,
$-(\eta_1 + \delta_1) \cdot z^*_{s_1} + \eta_1 \cdot z^*_{s_2} = -(\eta_1 + \delta_1) \cdot z^* + \eta_1 \cdot z^* = -\delta_1 \cdot z^*$. Hence, it is sufficient to compare the coefficient sum $\beta_{11}+\beta_{21}$ in the proposed model (\ref{eq:model2}) (Figure~\ref{fig:dataexample}(A)) to the negative of the interaction effect $-\delta_1$ in the Cox regression model (Figure~\ref{fig:dataexample}(B)). 

\clearpage

\section{Additional simulation results}
\label{app:additional_simulation}

\begin{figure}[h!]
    \centering
    \includegraphics[width = 0.9\linewidth]{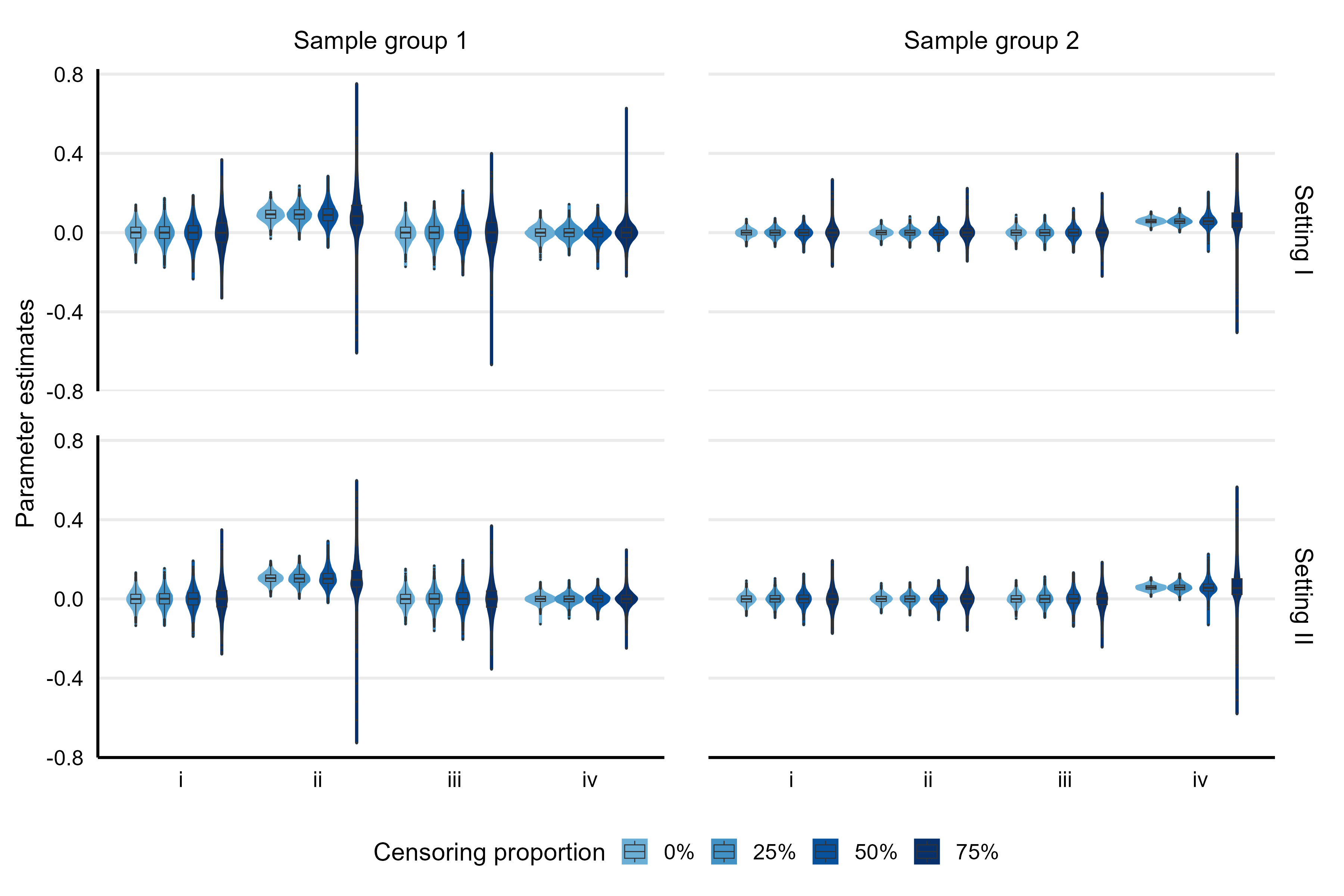}
    \caption{Results from the simulation study for sample sizes $n_1=150, n_2=300$. The figure presents boxplots of the estimates of $\beta_{j1}$, $j=1,2$, for different covariate space dimensions $p\in \{2,4\}$. Settings I and II refer to unequal and equal  Weibull shape parameters, respectively.}
    \label{fig:boxplots_betas_n150}
\end{figure}

\begin{figure}[h!]
    \centering
    \includegraphics[scale = 0.55]{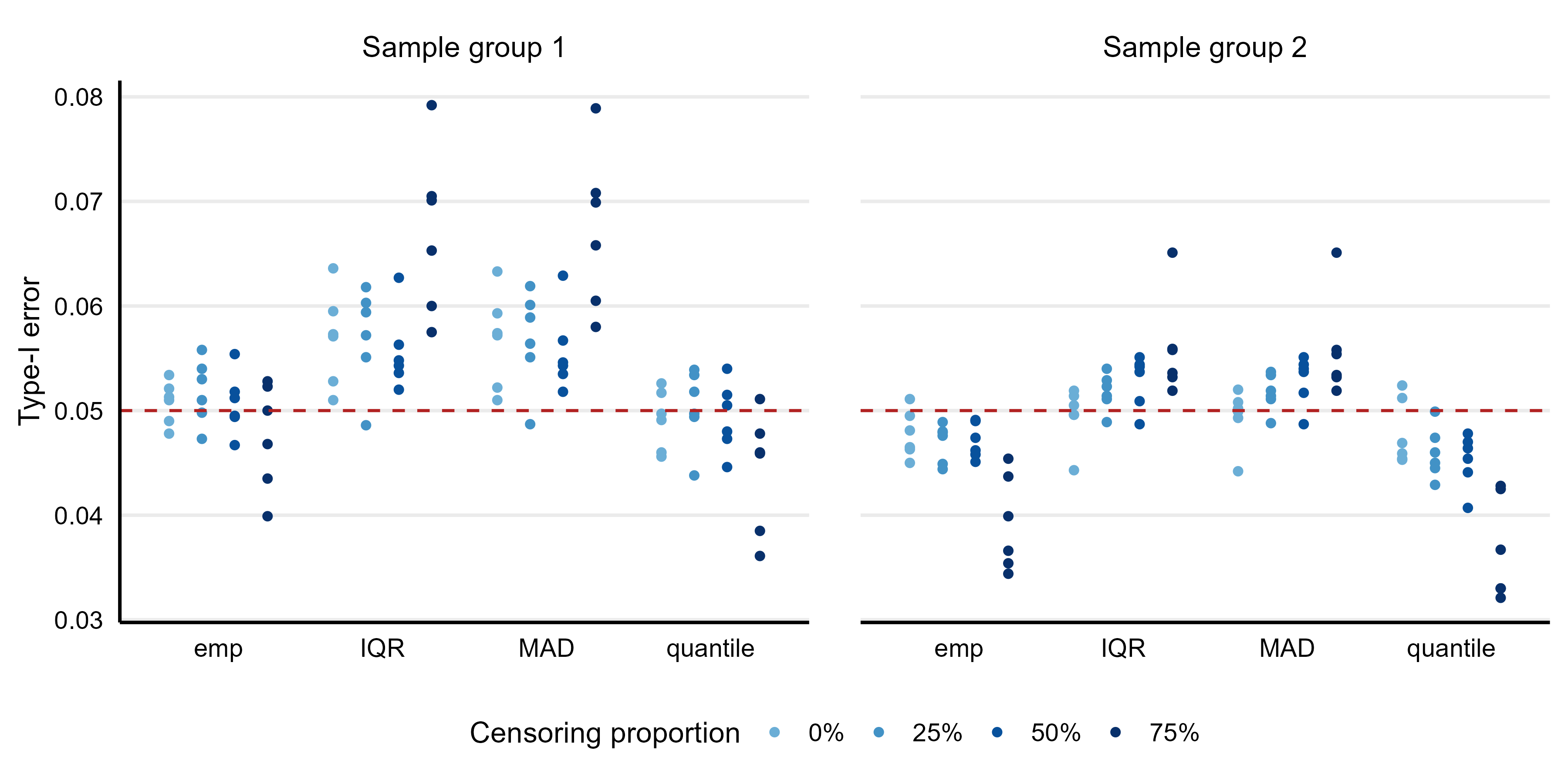}
    \caption{Type-I error rates of the tests for  $H_0^{(j)}:\beta_{j1}=0$ vs.\ $H_a^{(j)}: \beta_{j1}\neq 0$, $j=1,2$,  across all settings for sample sizes $n_1=150, n_2=300$. The nominal significance level $\alpha=5\%$ is displayed as a horizontal dashed line. The labels on the x-axes correspond to the four tests $\varphi_{j,\text{emp}}$, $\varphi_{j,\text{IQR}}$, $\varphi_{j,\text{MAD}}$ and $\varphi_{j,\text{quantile}}$.}
\label{fig:boxplots_type-I_power_n150}
\end{figure}

\begin{figure}[h!]
    \centering
    \includegraphics[width = 0.9\linewidth]{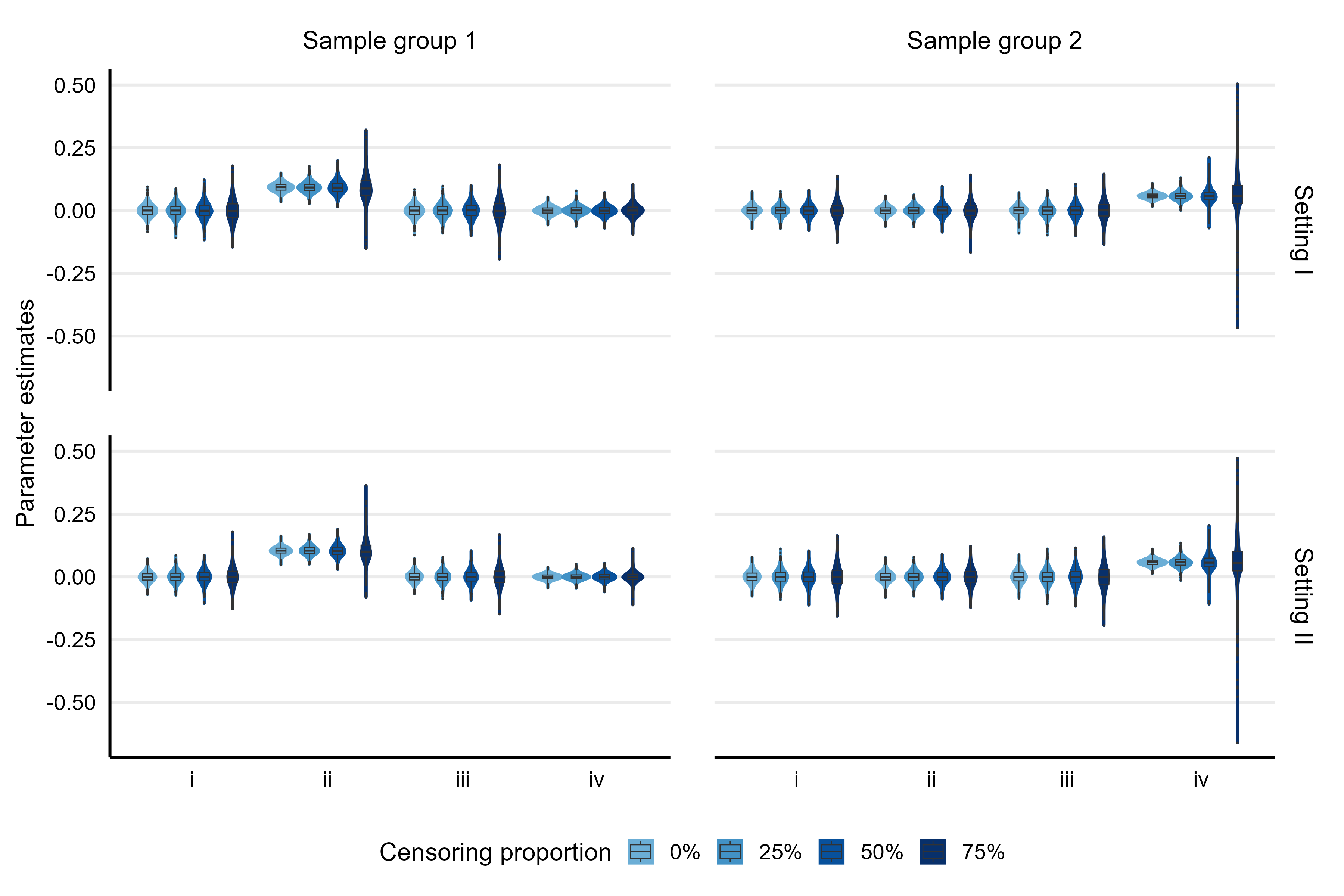}
    \caption{Results from the simulation study for sample sizes $n_1=500, n_2=300$. The figure presents boxplots of the estimates of $\beta_{j1}$, $j=1,2$, for different covariate space dimensions $p\in \{2,4\}$. Settings I and II refer to unequal and equal  Weibull shape parameters, respectively.}
    \label{fig:boxplots_betas_n300}
\end{figure}

\begin{figure}[h!]
    \centering
    \includegraphics[scale = 0.55]{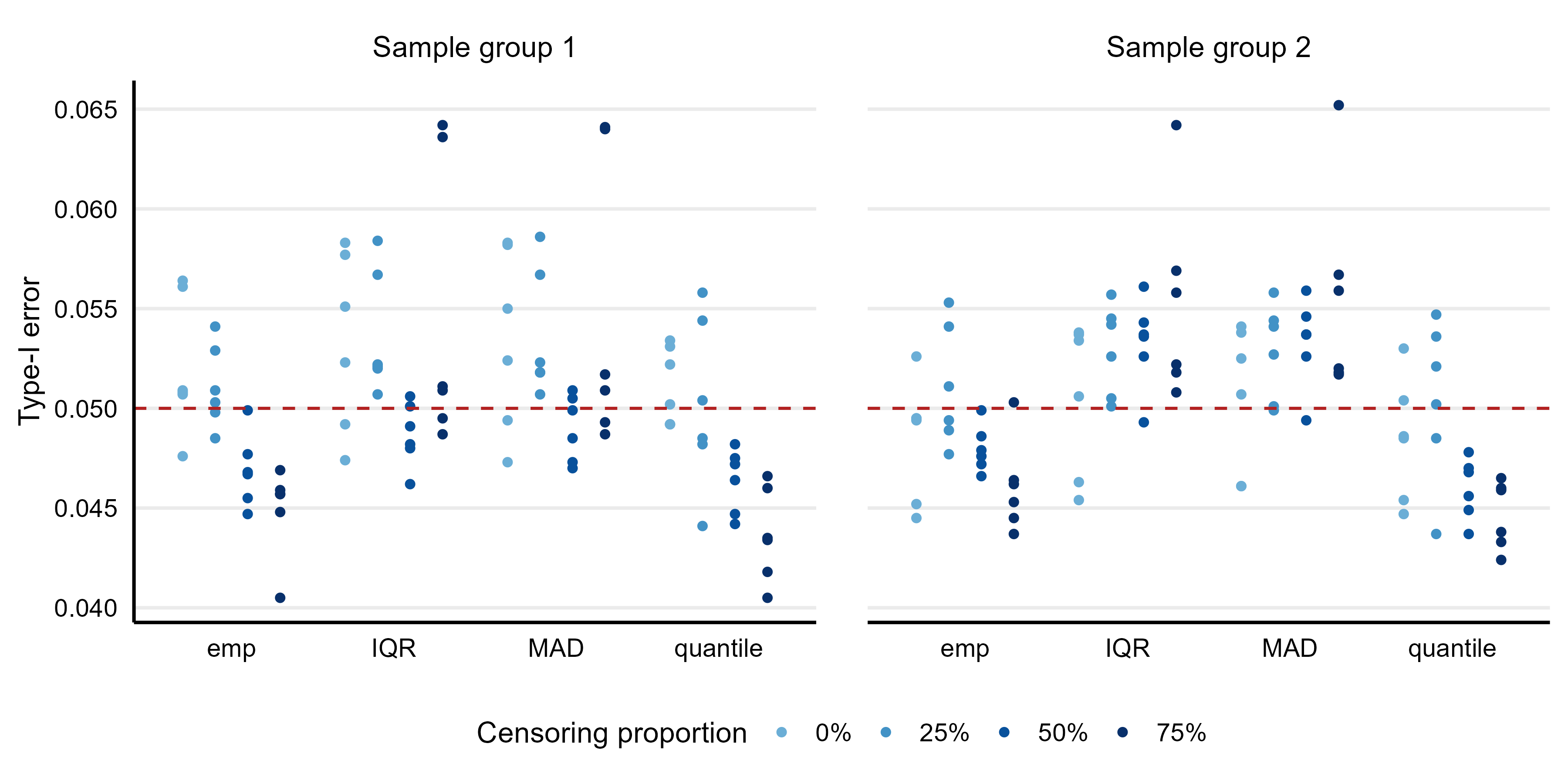}
    \caption{Type-I error rates of the tests for  $H_0^{(j)}:\beta_{j1}=0$ vs.\ $H_a^{(j)}: \beta_{j1}\neq 0$, $j=1,2$,  across all settings for sample sizes $n_1=500, n_2=300$. The nominal significance level $\alpha=5\%$ is displayed as a horizontal dashed line. The labels on the x-axes correspond to the four tests $\varphi_{j,\text{emp}}$, $\varphi_{j,\text{IQR}}$, $\varphi_{j,\text{MAD}}$ and $\varphi_{j,\text{quantile}}$.}
\label{fig:boxplots_type-I_power_n300}
\end{figure}

This appendix contains the simulation results for the sample sizes $(n_1,n_2) \in \{(150,300), (500, 300)\}$; see Tables~\ref{tab:rejection_p2} and~\ref{tab:rejection_p4} and Figures~\ref{fig:boxplots_betas_n150} to~\ref{fig:boxplots_type-I_power_n300}.
The results for $(n_1,n_2) =(150,300)$ are similar to those for $n_1=n_2=200$ which are included in the main part of the paper.
Increasing the sample sizes to $(n_1,n_2) =(500,300)$ generally reduced the variability of the parameter estimates, improved the type-I error control under $H_0^{(j)}$, and augmented the power of the tests under $H_1^{(j)}$, $j=1,2$.
One exception seems to be the case of $75\%$ censoring and $p_1=p_2=4$ where the power of the proposed tests seems to stagnate at rather low levels.
One reason for this could be that only the sample size $n_1$ is increased; indeed, the probability to reject $H_0^{(2)}$ stagnates whereas the probability to reject $H_0^{(1)}$ grows considerably.

\begin{table}[h!]
\begin{center}
\caption{Results from the simulation study. The table presents the estimated rejection rates (rounded to full percentage points) of the tests for $H_0^{(j)}:\beta_{j1}=0$ vs.\ $H_1^{(j)}: \beta_{j1}\neq 0$, $j=1,2$, and the corresponding $z$-tests in the Cox model for the presence of a group-covariate interaction (Scenarios i and ii,  $p_1=p_2=2$).
The underlined numbers indicate estimated powers under $H_1^{(j)}$.
Numbers are printed in bold when contained in the 95\%-Monte-Carlo error interval $[4.58\%,5.43\%]$ under $H_0^{(j)}$ or when representing the highest power under~$H_1^{(j)}$.
The abbreviations emp, IQR, MAD and quantile refer to the four tests $\varphi_{j,\text{emp}}$, $\varphi_{j,\text{IQR}}$, $\varphi_{j,\text{MAD}}$ and $\varphi_{j,\text{quantile}}$, respectively. \label{tab:rejection_p2}}
\vspace{2mm}
\scriptsize
\begin{tabular}{cccc|rrrr|r|rrrr|r}
  \hline
 &Weibull & $(\boldsymbol \gamma_1, \boldsymbol \gamma_2)$ & censoring & \multicolumn{5}{c|}{$H_0^{(1)}$ vs.\ $H_1^{(1)}$} & \multicolumn{5}{c}{$H_0^{(2)}$ vs.\ $H_1^{(2)}$} \\ 
 $(n_1,n_2)$ & shapes $(k_1,k_2)$ & scenario & proportion & emp & IQR & MAD & quantile & Cox & emp & IQR & MAD & quantile & Cox \\
  \hline
(150,300)  &(2,3)& i & 0 & \textbf{5} & \textbf{5} & \textbf{5} & \textbf{5} & 10 & \textbf{5} & \textbf{5} & \textbf{5} & \textbf{5} & 11 \\
   &(3,3)& i & 0 & \textbf{5} & 6 & 6 & \textbf{5} & 5 & \textbf{5} & 4 & 4 & \textbf{5} &  5 \\
   &(2,3)& ii & 0 & \underline{99} & \underline{99} & \underline{99} & \underline{99} & \underline{66} & \textbf{5} & \textbf{5} & \textbf{5} & {5} & {14} \\
   &(3,3)& ii & 0 & \underline{100} & \underline{100} & \underline{100} & \underline{100} & \underline{83} & \textbf{5} & \textbf{5} & \textbf{5} & \textbf{5} & 5 \\
   &(2,3)& i & 25 & 6 & 6 & 6 & \textbf{5} & 10 & \textbf{5} & \textbf{5} & \textbf{5} & 4 & 11 \\
   &(3,3)& i & 25 & \textbf{5} & 6 & 6 & \textbf{5} & 6 & \textbf{5} & \textbf{5} & \textbf{5} & \textbf{5} & 6 \\
   &(2,3)& ii & 25 & \underline{97} & \underline{97} & \underline{97} & \underline{96} & \underline{56} & 4 & \textbf{5} & \textbf{5} & 4 & 13 \\
   &(3,3)& ii & 25 & \underline{100} & \underline{100} & \underline{100} & \underline{100} & \underline{70} & \textbf{5} & \textbf{5} & \textbf{5} & \textbf{5} & \underline{5} \\
   &(2,3)& i & 50 & \textbf{5} & \textbf{5} & \textbf{5} & \textbf{5} & 10 & \textbf{5} & \textbf{5} & \textbf{5} & 4 & 11 \\
   &(3,3)& i & 50 & 6 & 6 & 6 & \textbf{5} & \textbf 5 & \textbf{5} & \textbf{5} & \textbf{5} & \textbf{5} & 5 \\
   &(2,3)& ii & 50 & \underline{82} & \underline{84} & \underline{84} & \underline{82} & \underline{41} & {5} & 6 & 6 & {5} & 12 \\
   &(3,3)& ii & 50 & \underline{97} & \underline{98} & \underline{98} & \underline{96} & \underline{54} & \textbf{5} & \textbf{5} & \textbf{5} & \textbf{5} & 6 \\
   &(2,3)& i & 75 & \textbf{5} & 7 & 7 & \textbf{5} & 10 & 3 & 6 & 6 & 3 & 10 \\
   &(3,3)& i & 75 & \textbf{5} & 6 & 6 & \textbf{5} & \textbf 5 & 4 & \textbf{5} & \textbf{5} & 4 & 5 \\
   &(2,3)& ii & 75 & \underline{21} & \underline{38} & \underline{38} & \underline{19} & \underline{25} & 4 & 7 & 7 & 4 & 11 \\
   &(3,3)& ii & 75 & \underline{36} & \underline{60} & \underline{60} & \underline{31} & \underline{30} & 4 & \textbf{5} & \textbf{5} & 3 & \textbf 5 \\ \hline \\[-.3cm]
(500,300) &(2,3)& i & 0 & \textbf{5} & \textbf{5} & \textbf{5} & \textbf{5} & 6 & 4 & {5} & \textbf{5} & 4 & 8 \\
   &(3,3)& i & 0 & 6 & 6 & 6 & \textbf{5} & 5 & \textbf{5} & \textbf{5} & \textbf{5} & \textbf{5} & 5 \\
   &(2,3)& ii & 0 & \underline{100} & \underline{100} & \underline{100} & \underline{100} & \underline{96} & \textbf{5} & \textbf{5} & \textbf{5} & \textbf{5} & 9 \\
   &(3,3)& ii & 0 & \underline{100} & \underline{100} & \underline{100} & \underline{100} & \underline{100} & \textbf{5} & \textbf{5} & \textbf{5} & \textbf{5} & 5 \\
   &(2,3)& i & 25 & \textbf{5} & \textbf{5} & \textbf{5} & \textbf{5} & 6 & \textbf{5} & \textbf{5} & \textbf{5} & \textbf{5} & 8 \\
   &(3,3)& i & 25 & \textbf{5} & \textbf{5} & \textbf{5} & \textbf{5} & 5 & 6 & {5} & \textbf{5} & {5} & 5 \\
   &(2,3)& ii & 25 & \underline{100} & \underline{100} & \underline{100} & \underline{100} & \underline{88} & \textbf{5} & 6 & 6 & \textbf{5} & 10 \\
   &(3,3)& ii & 25 & \underline{100} & \underline{100} & \underline{100} & \underline{100} & \underline{99} & \textbf{5} & \textbf{5} & {5} & \textbf{5} & 5 \\
   &(2,3)& i & 50 & \textbf{5} & \textbf{5} & \textbf{5} & \textbf{5} & 6 & \textbf{5} & \textbf{5} & \textbf{5} & \textbf{5} & 8 \\
   &(3,3)& i & 50 & \textbf{5} & \textbf{5} & \textbf{5} & \textbf{5} & 5 & \textbf{5} & \textbf{5} & \textbf{5} & \textbf{5} & 5 \\
   &(2,3)& ii & 50 & \underline{100} & \underline{100} & \underline{100} & \underline{100} & \underline{73} & \textbf{5} & \textbf{5} & \textbf{5} & {5} & 9 \\
   &(3,3)& ii & 50 & \underline{100} & \underline{100} & \underline{100} & \underline{100} & \underline{93} & \textbf{5} & \textbf{5} & \textbf{5} & 4 & 5 \\
   &(2,3)& i & 75 & 4 & \textbf{5} & \textbf{5} & 4 & 6 & 4 & \textbf{5} & \textbf{5} & 4 & 7 \\
   &(3,3)& i & 75 & {5} & \textbf{5} & \textbf{5} & 4 & 5 & \textbf{5} & {5} & \textbf{5} & 4 & 5 \\
   &(2,3)& ii & 75 & \underline{77} & \underline{85} & \underline{86} & \underline{72} & \underline{45} & \textbf{5} & 6 & 6 & \textbf{5} & 8 \\
   &(3,3)& ii & 75 & \underline{98} & \underline{99} & \underline{99} & \underline{97} & \underline{69} & \textbf{5} & \textbf{5} & \textbf{5} & \textbf{5} & 5 \\
   \hline
\end{tabular}
\end{center}
\end{table}

\begin{table}[h!]
\begin{center}
    \caption{Results from the simulation study. The table presents the estimated rejection rates (rounded to full percentage points) of the tests for $H_0^{(j)}:\beta_{j1}=0$ vs.\ $H_1^{(j)}: \beta_{j1}\neq 0$, $j=1,2$, and the corresponding $z$-tests in the Cox model for the presence of a group-covariate interaction (Scenarios iii and iv,  $p_1=p_2=4$).
The underlined numbers indicate estimated powers under $H_1^{(j)}$.
Numbers are printed in bold when contained in the 95\%-Monte-Carlo error interval $[4.58\%,5.43\%]$ under $H_0^{(j)}$ or when representing the highest power under~$H_1^{(j)}$.
The abbreviations emp, IQR, MAD and quantile refer to the four tests $\varphi_{j,\text{emp}}$, $\varphi_{j,\text{IQR}}$, $\varphi_{j,\text{MAD}}$ and $\varphi_{j,\text{quantile}}$, respectively. \label{tab:rejection_p4}}
\scriptsize
\vspace{2mm}
\begin{tabular}{cccc|rrrr|r|rrrr|r}
  \hline
 &Weibull & $(\boldsymbol \gamma_1, \boldsymbol \gamma_2)$ & censoring & \multicolumn{5}{c|}{$H_0^{(1)}$ vs.\ $H_1^{(1)}$} & \multicolumn{5}{c}{$H_0^{(2)}$ vs.\ $H_1^{(2)}$} \\ 
 $(n_1,n_2)$ & shapes $(k_1,k_2)$ & scenario & proportion & emp & IQR & MAD & quantile & Cox & emp & IQR & MAD & quantile & Cox \\
  \hline
(150,300) &(2,3)& iii & 0 & \textbf{5} & \textbf{5} & \textbf{5} & {5} & 10 & \textbf{5} & \textbf{5} & \textbf{5} & \textbf{5} & 11 \\
   &(3,3)& iii & 0 & \textbf{5} & 6 & 6 & \textbf{5} & 5 & \textbf{5} & \textbf{5} & \textbf{5} & {5} & 5 \\
   &(2,3)& iv & 0 & \textbf{5} & 6 & 6 & \textbf{5} & 10 & \underline{100} & \underline{100} & \underline{100} & \underline{100} & \underline{100} \\
   &(3,3)& iv & 0 & \textbf{5} & 6 & 6 & \textbf{5} & 5 & \underline{100} & \underline{100} & \underline{100} & \underline{100} & \underline{100} \\
   &(2,3)& iii & 25 & \textbf{5} & \textbf{5} & \textbf{5} & 4 & 10 & 4 & \textbf{5} & \textbf{5} & \textbf{5} & 11 \\
   &(3,3)& iii & 25 & \textbf{5} & 6 & 6 & \textbf{5} & 6 & \textbf{5} & \textbf{5} & \textbf{5} & \textbf{5} & 5 \\
   &(2,3)& iv & 25 & \textbf{5} & 6 & 6 & \textbf{5} & 10 & \underline{100} & \underline{100} & \underline{100} & \underline{100} & \underline{98} \\
   &(3,3)& iv & 25 & \textbf{5} & 6 & 6 & \textbf{5} & 5 & \underline{100} & \underline{100} & \underline{100} & \underline{100} & \underline{100} \\
   &(2,3)& iii & 50 & \textbf{5} & {5} & {5} & \textbf{5} & 10 & \textbf{5} & \textbf{5} & \textbf{5} & 4 & 12 \\
   &(3,3)& iii & 50 & \textbf{5} & \textbf{5} & \textbf{5} & \textbf{5} & 6 & \textbf{5} & {5} & {5} & \textbf{5} & 6 \\
   &(2,3)& iv & 50 & \textbf{5} & 6 & 6 & \textbf{5} & 10 & \underline{87} & \underline{92} & \underline{92} & \underline{83} & \underline{91} \\
   &(3,3)& iv & 50 & \textbf{5} & \textbf{5} & \textbf{5} & 4 & 5 & \underline{81} & \underline{86} & \underline{86} & \underline{77} & \underline{100} \\
   &(2,3)& iii & 75 & \textbf{5} & 7 & 7 & \textbf{5} & 9 & 4 & 6 & 6 & 3 & 11 \\
   &(3,3)& iii & 75 & 4 & 6 & 6 & 4 & 6 & {5} & \textbf{5} & \textbf{5} & 4 & 6 \\
   &(2,3)& iv & 75 & \textbf{5} & 8 & 8 & \textbf{5} & 9 & \underline{16} & \underline{34} & \underline{34} & \underline{12} & \underline{65} \\
   &(3,3)& iv & 75 & 4 & 7 & 7 & 4 & 6 & \underline{13} & \underline{31} & \underline{31} & \underline{9} & \underline{89} \\ \hline \\[-.3cm]
  (500,300) &(2,3)& iii & 0 & \textbf{5} & \textbf{5} & \textbf{5} & \textbf{5} & 5 & {5} & \textbf{5} & \textbf{5} & {5} & 8 \\
   &(3,3)& iii & 0 & \textbf{5} & \textbf{5} & \textbf{5} & \textbf{5} & 5 & \textbf{5} & \textbf{5} & \textbf{5} & \textbf{5} & 5 \\
   &(2,3)& iv & 0 & 6 & 6 & 6 & \textbf{5} & 8 & \underline{100} & \underline{100} & \underline{100} & \underline{100} & \underline{100} \\
   &(3,3)& iv & 0 & \textbf{5} & 6 & 6 & \textbf{5} & 5 & \underline{100} & \underline{100} & \underline{100} & \underline{100} & \underline{100} \\
   &(2,3)& iii & 25 & \textbf{5} & \textbf{5} & \textbf{5} & 5 & 5 & \textbf{5} & \textbf{5} & \textbf{5} & \textbf{5} & 8 \\
   &(3,3)& iii & 25 & \textbf{5} & 6 & 6 & 4 & 5 & \textbf{5} & \textbf{5} & \textbf{5} & 4 & 5 \\
   &(2,3)& iv & 25 & \textbf{5} & 6 & 6 & 6 & 8 & \underline{100} & \underline{100} & \underline{100} & \underline{100} & \underline{100} \\
   &(3,3)& iv & 25 & \textbf{5} & \textbf{5} & \textbf{5} & \textbf{5} & 6 & \underline{100} & \underline{100} & \underline{100} & \underline{100} & \underline{100} \\
   &(2,3)& iii & 50 & \textbf{5} & \textbf{5} & \textbf{5} & \textbf{5} & \textbf 5 & \textbf{5} & 6 & 6 & \textbf{5} & 8 \\
   &(3,3)& iii & 50 & \textbf{5} & \textbf{5} & \textbf{5} & \textbf{5} & 5 & \textbf{5} & \textbf{5} & {5} & 4 & 6 \\
   &(2,3)& iv & 50 & 4 & \textbf{5} & \textbf{5} & 4 & 8 & \underline{91} & \underline{95} & \underline{95} & \underline{88} & \underline{100} \\
   &(3,3)& iv & 50 & {5} & \textbf{5} & \textbf{5} & 4 & 5 & \underline{83} & \underline{89} & \underline{89} & \underline{79} & \underline{100} \\
   &(2,3)& iii & 75 & \textbf{5} & \textbf{5} & \textbf{5} & \textbf{5} & 6 & 4 & 6 & 6 & 4 & 7 \\
   &(3,3)& iii & 75 & \textbf{5} & \textbf{5} & \textbf{5} & \textbf{5} & 6 & \textbf{5} & 6 & 7 & \textbf{5} & 6 \\
   &(2,3)& iv & 75 & {5} & 6 & 6 & 4 & 8 & \underline{16} & \underline{34} & \underline{34} & \underline{13} & \underline{99} \\
   &(3,3)& iv & 75 & 4 & 6 & 6 & 4 & 6 & \underline{12} & \underline{30} & \underline{30} & \underline{8} & \underline{100} \\
   \hline
\end{tabular}
\end{center}
\end{table}

\newpage
\phantom{X}
\newpage
\phantom{X}
\newpage

\newpage
\phantom{X}
\newpage

\section{Supplementary results from the analysis of the SUCCESS-A study data}
\label{app:success}

\begin{table}[h!]
\centering
\caption{Descriptive summary of the variables of the SUCCESS-A study data (SD = standard deviation).\\}
\label{tab:supplement_SUCCESS_A} 
\begin{tabular}{@{}p{5.5cm}>{\centering\arraybackslash}p{2.8cm}>{\centering\arraybackslash}p{3cm}>{\centering\arraybackslash}p{3cm}@{}}
  \hline
  \textbf{Characteristic} & \textbf{Total \ \ (N=3,652)} & \textbf{Control (N=1,840)}  &  \textbf{Intervention (N=1,812)}  \\ 
  \hline
     \textbf{Age (years)} &  & & \\ 
     \hspace{0.4cm} Mean (SD) & 53.5 (10.5) & 53.9 (10.4) & 53.1 (10.6) \\ 
     \hspace{0.4cm} Median [IQR] & 53.0 [46.0; 62.0]  & 54.0 [46.0; 62.0]  & 53.0 [45.0; 62.0] \\ 
     \hline
     \textbf{BMI ($kg/m^2$)} &  \\ 
     \hspace{0.4cm} Mean (SD) & 26.3 (5.0) & 26.3 (5.1) & 26.2 (5.0)  \\ 
     \hspace{0.4cm} Median [IQR] & 25.4 [22.6; 29.1]  & 25.3 [22.6; 29.3]  & 25.5 [22.7; 29.0]\\ 
\hline
     \textbf{Tumor stage} &  \\ 
     \hspace{0.4cm} pT1 & 1,516 (41.5\%) & 752 (40.9\%) & 764 (42.2\%)\\ 
     \hspace{0.4cm} pT2 & 1,888 (51.7\%) & 954 (51.8\%) & 934 (51.5\%)\\ 
     \hspace{0.4cm} pT3 & 196 (5.4\%) & 108 (5.9\%) & 88 (4.9\%)\\ 
     \hspace{0.4cm} pT4 & 52 (1.4\%) & 26 (1.4\%)  & 26 (1.4\%)\\ 
\hline
     \textbf{Tumor grade} &  \\ 
     \hspace{0.4cm} G1 & 173 (4.7\%)  & 77 (4.2\%)  & 96 (5.3\%) \\ 
     \hspace{0.4cm} G2 & 1,746 (47.8\%) & 876 (47.6\%) & 870 (48.0\%) \\ 
     \hspace{0.4cm} G3 & 1,733 (47.5\%) & 887 (48.2\%) & 846 (46.7\%)\\ 
\hline
     \textbf{Lymph node status} &  \\ 
     \hspace{0.4cm} pN+ & 2,401 (65.7\%) & 1,217 (66.1\%) & 1,184 (65.3\%)\\ 
     \hspace{0.4cm} pN0 & 1,251 (34.3\%) & 623 (33.9\%)  & 628 (34.7\%)  \\ 
\hline
     \textbf{Tumor type} &  \\ 
     \hspace{0.4cm} ductal & 2,992 (81.9\%) & 1,507 (81.9\%)  & 1,485 (82.0\%)\\ 
     \hspace{0.4cm} lobular & 409 (11.2\%) & 207 (11.3\%)  & 202 (11.1\%)\\ 
     \hspace{0.4cm} other & 251 (6.9\%) & 126 (6.8\%)  & 125 (6.9\%)\\ 
\hline
     \textbf{Estrogen receptor status} &  \\ 
     \hspace{0.4cm} ER- & 1,229 (33.7\%) & 601 (32.7\%) & 628 (34.7\%) \\ 
     \hspace{0.4cm} ER+ & 2,423 (66.3\%) & 1,239 (67.3\%) & 1,184 (65.3\%)\\ 
\hline
     \textbf{Progesterone receptor status} &  \\ 
     \hspace{0.4cm} PR- & 1,494 (40.9\%) & 753 (40.9\%) & 741 (40.9\%) \\ 
     \hspace{0.4cm} PR+ & 2,158 (59.1\%) & 1,087 (59.1\%) & 1,071 (59.1\%) \\ 
\hline
     \textbf{HER2 status} &  \\ 
     \hspace{0.4cm} HER2- & 2,772 (75.9\%) & 1,402 (76.2\%) & 1,370 (75.6\%) \\ 
     \hspace{0.4cm} HER2+ & 880 (24.1\%) & 438 (23.8\%) & 442 (24.4\%)  \\ 
\hline
     \textbf{Menopausal status} &  \\ 
     \hspace{0.4cm} pre & 1,524 (41.7\%) & 754 (41.0\%) & 770 (42.5\%)\\ 
     \hspace{0.4cm} post & 2,128 (58.3\%) & 1,086 (59.0\%) & 1,042 (57.5\%) \\ 
   \hline
\end{tabular}
\end{table}

\begin{table}[]
\centering
\caption{Analysis of the SUCCESS-A study data. The table presents a descriptive summary of the subgroups defined by benefit from control and benefit from the intervention (as defined in Section \ref{sec:data}).\\}
\label{tab:supplement_SUCCESS_A_benefit_harm} 
\begin{tabular}[h]{p{5.5cm}>{\centering\arraybackslash}p{2.8cm}>{\centering\arraybackslash}p{3cm}} 
\hline
  \textbf{Characteristic} & \textbf{Control better (N=1,112)}  &  \textbf{Intervention better (N=1,311)}  \\ 
\hline
     \textbf{Age (years)}  & & \\ 
     \hspace{0.4cm} Mean (SD) & 55.0 (10.0)  & 52.0 (11.0) \\ 
     \hspace{0.4cm} Median [IQR] &  56.0 [48.0; 63.0]  & 51.0 [44.0; 61.0] \\ 
\hline
     \textbf{BMI ($kg/m^2$)} &  \\ 
     \hspace{0.4cm} Mean (SD) & 26.2 (4.9)  & 26.1 (4.8)  \\ 
     \hspace{0.4cm} Median [IQR]   & 25.3 [22.7; 29.1]  & 25.3 [22.6; 29.0]\\ 
\hline
     \textbf{Tumor stage} &  \\ 
     \hspace{0.4cm} pT1 & 658 (59.2\%) & 313 (23.9\%)\\ 
     \hspace{0.4cm} pT2 & 441 (39.7\%) & 857 (65.4\%) \\
     \hspace{0.4cm} pT3 & 13 (1.2\%) & 108 (8.2\%) \\
     \hspace{0.4cm} pT4 & 0 (0.0\%) & 33 (2.5\%) \\
\hline
     \textbf{Tumor grade} &  \\ 
     \hspace{0.4cm} G1 & 156 (14.0\%)  &  0 (0.0\%) \\ 
     \hspace{0.4cm} G2 & 653 (58.7\%) & 345 (26.3\%) \\ 
     \hspace{0.4cm} G3 & 303 (27.2\%) & 966 (73.7\%) \\ 
\hline
     \textbf{Lymph node status} &  \\ 
     \hspace{0.4cm} pN+ & 482 (43.3\%) & 1077 (82.2\%)\\ 
     \hspace{0.4cm} pN0 & 630 (56.7\%) & 234 (17.8\%) \\ 
\hline
     \textbf{Tumor type} &  \\ 
     \hspace{0.4cm} ductal  & 875 (78.7\%) & 1,189 (90.7\%) \\ 
     \hspace{0.4cm} lobular & 130 (11.7\%) & 92 (7.0\%) \\ 
     \hspace{0.4cm} other   & 107 (9.6\%)  & 30 (2.3\%) \\
\hline
     \textbf{Estrogen receptor status} &  \\ 
     \hspace{0.4cm} ER$-$ & 137 (12.3\%) & 722 (55.1\%) \\ 
     \hspace{0.4cm} ER+   & 975 (87.7\%) & 589 (44.9\%) \\ 
\hline
     \textbf{Progesterone receptor status} &  \\ 
     \hspace{0.4cm} PR$-$ & 193 (17.4\%) & 872 (66.5\%) \\ 
     \hspace{0.4cm} PR+   & 919 (82.6\%) & 439 (33.5\%) \\ 
\hline
     \textbf{HER2 status} &  \\ 
     \hspace{0.4cm} HER2$-$ & 764 (68.7\%) & 1,050 (80.1\%) \\ 
     \hspace{0.4cm} HER2+   & 348 (31.3\%) & 261 (19.9\%) \\ 
\hline
     \textbf{Menopausal status} &  \\ 
     \hspace{0.4cm} pre  & 340 (30.6\%) & 666 (50.8\%)\\ 
     \hspace{0.4cm} post & 772 (69.4\%) & 645 (49.2\%) \\ 
\hline
     \textbf{Molecular tumor subtype} &  \\ 
     \hspace{0.4cm} HER2 positive    & 348 (31.3\%) & 261 (19.9\%) \\ 
     \hspace{0.4cm} Luminal A-like	& 591 (53.1\%) & 233 (17.8\%) \\ 
     \hspace{0.4cm} Luminal B-like	& 138 (12.4\%) & 317 (24.2\%) \\ 
     \hspace{0.4cm} Triple negative	& 35 (3.1\%) & 500 (38.1\%) \\ 
\hline
\end{tabular}
\end{table}

\begin{figure}[t]
    \centering
    \hspace*{-1.5cm}
    \includegraphics[scale = 0.52]{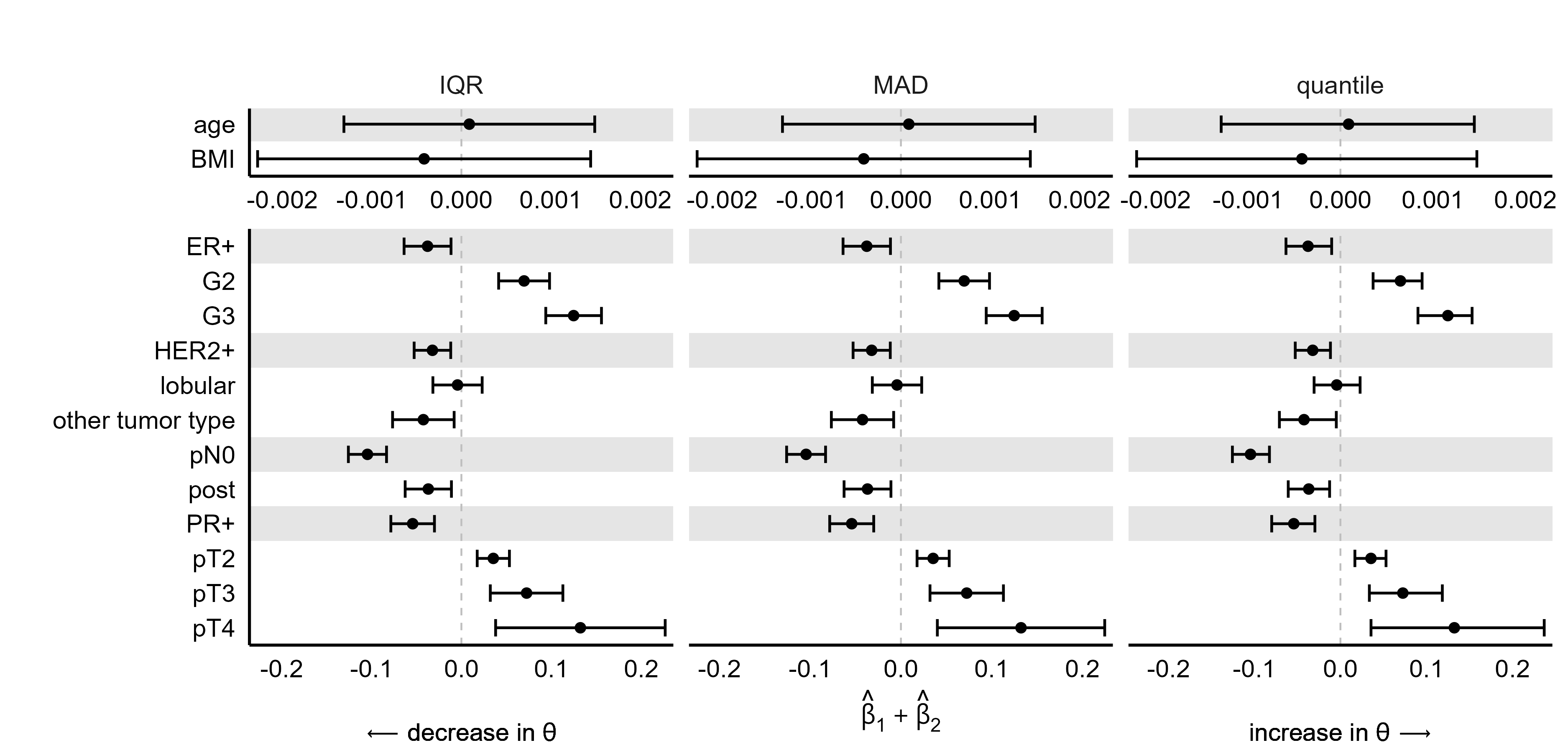}
    \caption{Analysis of the SUCCESS-A study data. The figure presents the estimated coefficients $\hat{\bs\beta}_1 + \hat{\bs\beta}_2 $ with 95\% confidence intervals, as obtained from the IQR, MAD, and empirical quantile methods. 
Positive values of $\hat{\bs\beta}_1 + \hat{\bs\beta}_2$ indicate a higher estimated probability of $\min(T_1, \tau) > \min(T_2, \tau)$, corresponding to a higher benefit of the intervention compared to the reference categories of the covariates. Age and BMI are measured in years and $kg/m^2$, respectively. The reference categories are ER$-$, G1, HER2$-$, ductal, pN+, pre-menopausal, PR$-$ and pT1. 
The estimated values of the intercept $\beta_0$ are 0.1560 [0.0783, 0.2336] for the IQR method, 0.1560 [0.788, 0.2332] for the MAD method, and 0.1560 [0.8441, 0.2313] for the empirical quantile method.}
    \label{fig:supp_coefficients}
\end{figure}

\begin{figure}[t]
    \centering
    \includegraphics[width = \textwidth]{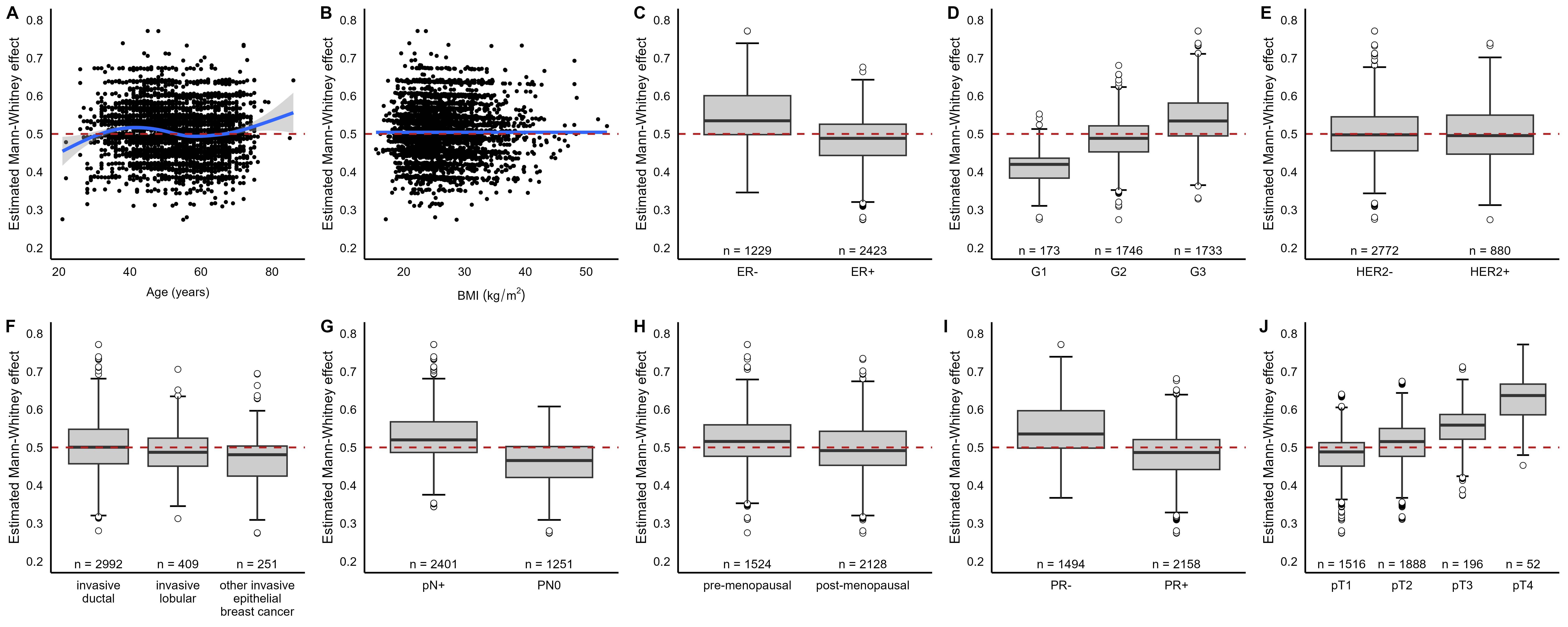}
    \caption{Analysis of the SUCCESS-A study data. The figures present the estimated tie-corrected truncated Mann-Whitney effects of the study participants ($n=$ 3,652) derived from \eqref{eq:predictions} and stratified by covariates. The blue lines in (A) and (B) were obtained using LOESS smoothing.}
    \label{fig:supp_predictions}
\end{figure}

\end{document}